# Temperature-dependent optical properties of gold thin films

Harsha Reddy[1,#], Urcan Guler[2,#,*], Alexander V Kildishev[1,2], Alexandra Boltasseva[1,2], Vladimir M. Shalaev[1,2,*]

[1]School of Electrical & Computer Engineering and Birck Nanotechnology Center, Purdue University, West Lafayette, IN 47907, USA

[2]Nano-Meta Technologies, Inc., 1281 Win Hentschel Blvd, West Lafayette, IN 47906, USA

**ABSTRACT:** Understanding the temperature dependence of the optical properties of thin metal films is critical for designing practical devices for high temperature applications in a variety of research areas, including plasmonics and near-field radiative heat transfer. Even though the optical properties of bulk metals at elevated temperatures have been studied, the temperature-dependent data for thin metal films, with thicknesses ranging from few tens to few hundreds of nanometers, is largely missing. In this work we report on the optical constants of single- and polycrystalline gold thin films at elevated temperatures in the wavelength range from 370 to 2000 nm. Our results show that while the real part of the dielectric function changes marginally with increasing temperature, the imaginary part changes drastically. For 200-nm-thick single- and polycrystalline gold films the imaginary part of the dielectric function at 500 $^{0}$C becomes nearly twice larger than that at room temperature. In contrast, in thinner films (50-nm and 30-nm) the imaginary part can show either increasing or decreasing behavior within the same temperature range and eventually at 500 $^{0}$C it becomes nearly 3-4 times larger than that at room temperature. The increase in the imaginary part at elevated temperatures significantly reduces the surface plasmon polariton propagation length and the quality factor of the localized surface plasmon resonance for a spherical particle. We provide experiment-fitted models to describe the temperature-dependent gold dielectric function as a sum of one Drude and two critical point oscillators. These causal analytical models could enable accurate multiphysics modelling of gold-based nanophotonic and plasmonic elements in both frequency and time domains.

**KEYWORDS:** *nanophotonics, plasmonics, metamaterials, metal optics, high temperatures, gold thin films, analytical models*

Nanometer-scale field localization is at the heart of metal-based nanophotonics, namely plasmonics[1-4] . In plasmonic nanostructures, strong field confinement - or so-called 'hot spots' - arise due to the excitation of subwavelength oscillations of free electrons coupled to the incident electromagnetic field at the metal-dielectric interface, known as surface plasmons[5]. Such nanoscale hot spots lead to high energy densities that inevitably increase the local temperature of the plasmonic material under study. Recently, there has been a growing interest in plasmonics-based local heating applications such as heat-assisted magnetic recording, thermophotovoltaics, and photothermal therapy[6,7]. However, theoretical modeling of plasmonic structures in such local-heating based systems has so far been performed using room-temperature optical constants, i.e. with thermal analysis and optical material properties being decoupled. Therefore, probing the temperature dependence of the optical properties of thin metal films is critical for both gaining an insight into the physical process associated with elevated temperatures and for accurate modeling of devices for high-temperature applications. Incorporating temperature dependence into causal experiment-fitted material models would be crucial for time-domain numerical studies of plasmonic elements[8], such as for spasers[9], plasmonic nanolasers[10,11] and plasmon-assisted photocatalysis[12,13] .

The optical properties of bulk metals at elevated temperatures have been studied previously[14,15]. However, as pointed out by Sundari *et al*[16] these studies only report on the imaginary part of the dielectric constant. More recently, there have been reports on the temperature dependent optical constants of 200-nm-thick silver films[16-18] as well as the temperature changes in the surface plasmon resonance of gold nanoparticles embedded in silica at elevated temperatures[19]. At low temperatures, optical properties of ultrathin Au films in the mid- and far-infrared regions[20] and the optical response of gold nanorods and plasmonic crystals[21] have also been studied. However, a comprehensive study of the optical properties of gold thin films with varying thicknesses over a wide wavelength range at elevated temperatures has not been conducted. Here we report on the temperature dependence of the optical properties of gold thin films of different thicknesses and different crystallinities. Specifically, we measured the optical constants of 200-nm, 50-nm and 30-nm-thick polycrystalline (PC) and 200-nm-thick single crystalline (SC) samples in the wavelength range from 370 to 2000 nm (sample preparation details are available in the Methods section). The 200-nm (both PC and SC) and 50-nm-thick films were probed at temperatures up to 500 $^0$C and the 30-nm-thick films were heated to temperatures up to 450 $^0$C. The surface morphology of the 30-nm-thick samples was significantly damaged when heated to 450 $^0$C. Hence they were not probed till 500 $^0$C. Further, the 200-nm-thick samples were subjected to multiple heating cycles and the changes in their optical properties over repeated heating were monitored. The temperature dependent measurements were enabled by integrating a heating stage into our Variable Angle Spectroscopic Ellipsometer (VASE) setup.

## EXPERIMENTAL SETUP AND MODELING APPROACH

A heating stage (Linkam Scientific Model TS1500) was mounted onto our VASE setup in order to probe the optical properties of the samples at high temperatures. The stage had the capability of heating the sample to temperatures up to 1500 $^0$C and a rated temperature stability of $\pm$ 2 $^0$C. In order to prevent rapid thermal expansion, the samples were heated and cooled down at a rate of 3 $^0$C/min. At temperatures above 450 $^0$C the noise due to background thermal emission saturates the detector. In order to reduce the background thermal emission from reaching the VASE detector a pinhole was introduced into the beam path, similar to that used in an earlier paper[22] (see supporting information for a more detailed discussion on the experimental set up). This enabled accurate measurements for temperatures over 450 $^0$C.

The VASE data were then fitted with a Drude and two Critical Point (DCP)[23,24] model, using the commercial software WVASE32, to extract the real ( $\varepsilon_1$ ) and imaginary ( $\varepsilon_2$ ) parts of the complex dielectric function ( $\hat{\varepsilon}(\omega) = \varepsilon_1 + i\varepsilon_2$ ). We use the following form of the DCP model (Eqn (1.1)).

$$\hat{\varepsilon}(\omega) = \varepsilon_\infty - \frac{\omega_p^2}{\omega^2 + i\Gamma_D\omega} + \sum_{j=1}^{2} C_j\Omega_j\left(\frac{e^{i\phi_j}}{\Omega_j - \omega - i\gamma_j} - \frac{e^{-i\phi_j}}{\Omega_j + \omega + i\gamma_j}\right) \tag{1.1}$$

where $\varepsilon_\infty$, $\omega_{pu}$ and $\Gamma_D$ are the background dielectric constant, plasma frequency and Drude broadening, respectively. Furthermore, $C_j$, $\Omega_j$, $\gamma_j$ and $\phi_j$ are the oscillator strength, oscillator energy, oscillator damping and oscillator phase, respectively.

In general, any dielectric function in the frequency domain could be approximated using an [m/n] Pade approximant of an argument $-i\omega$[25]

$$\hat{\varepsilon}(\omega) \approx \frac{\alpha_0 + (-i\omega)\alpha_1 + ... + (-i\omega)^p \alpha_p + ... + (-i\omega)^m \alpha_m}{\beta_0 + (-i\omega)\beta_1 ... + (-i\omega)^q \beta_q + ... + ... + (-i\omega)^n} \tag{1.2}$$

with $\alpha_p, \beta_q \in \mathbb{R}$. By using the fundamental theorem of algebra, we may split (1.2) into a constant, a detached zero-pole, and a number ( $j_1$ ) of [0/1]-order terms, along with a number ( $j_2$ ) of [1/2]-order terms (where the case of multiple poles is omitted)

$$\hat{\varepsilon}(\omega) = \varepsilon_\infty - \frac{\sigma}{i\omega\varepsilon_0} + \sum_{j \in I_1} \frac{a_{0,j}}{b_{0,j} - i\omega} + \sum_{j \in I_2} \frac{a_{0,j} - i\omega a_{1,j}}{b_{0,j} - i\omega b_{1,j} - \omega^2} \tag{1.3}$$

here $\varepsilon_0$ is the electric permittivity of vacuum, $I_1 = \{I_1 \subset \mathbb{N} \mid 1 \le j \le j_2\}$ and $I_2 = \{I_2 \subset \mathbb{N} \mid j_1 + 1 \le j \le j_1 + j_2\}$ are non-overlapping ranges of indices. While initially $\varepsilon_\infty$ can be interpreted as the high-frequency approximant and $\sigma$ as a conductivity term, it can be shown that the universal approximation in frequency and time domains can be achieved with expression (1.3) and its time-domain analog in particular represents any set of the classical Debye, Drude, Lorentz, Sellmeier, and critical points terms[25].

Although any arbitrary $\hat{\varepsilon}(\omega)$ can be described using several additional oscillators they do not provide any physical insight. On the other hand, a DCP model describes the $\hat{\varepsilon}(\omega)$ with a minimal number of free parameters; it is causal[23] and is broadly used for modeling gold plasmonic elements in time-domain[26,8]. We therefore used a DCP model instead of a Lorentz oscillator model for fitting our experimental data.

While fitting the VASE data, the oscillator phases were kept fixed at $-\pi / 4$ and all other terms were supplied as fits. The fits for the 200-nm-thick films were obtained by treating the Au film as a semi-infinite layer. For the 50-nm and 30-nm-thick films a two layer model consisting of optical constants of glass and DCP model was used to extract the optical constants. Unlike the thicker films, whose optical constants are independent of the thickness parameter, the extracted optical constants of thin films are highly sensitive to the supplied thickness parameter while fitting the VASE data. During deposition, the thicknesses of the PC films were monitored using a crystal oscillator. However, given the sensitivity of the optical constants on the thickness parameter, the precise value of the thickness was needed in order to extract the accurate values of the optical constants. Therefore, we verified the thicknesses using SEM cross-sectional imaging and found them to be consistent with those measured using the crystal oscillator. For thinner films, in addition to the oscillator phases we keep the thickness fixed at 50 nm and 30 nm thick respectively while the other terms in the model were supplied as fit parameters. The Mean Square Error's (MSE) for all the measurements were less than 2 indicating that the fits were good (the temperature dependent DCP models are shown in Tables S1-S7 in the supporting information).

## EXPERIMENTAL RESULTS AND DISCUSSION

The experimentally obtained complex dielectric permittivity of the 200-nm thick PC film as a function of wavelength from room temperature to 500 $^0$C for three heating cycles are shown in Figure 1. The imaginary part of the dielectric function $\varepsilon_2$ (Figure 1b,d,f) increases monotonically with increasing temperature for all heating cycles. However, for the first cycle the increase in $\varepsilon_2$ (Figure 1b) is not uniform which is due to the annealing effects and grain movements that occur at high temperatures[27]. The increase in $\varepsilon_2$ is more uniform for the subsequent cycles (Figure 1d,f). At longer wavelengths ($\lambda > 900nm$), where the inter-band transitions are insignificant, the imaginary part at 500 $^0$C becomes nearly twice as large as it is at room temperature. This behavior in $\varepsilon_2$ can be understood by noting that the scattering rates of the free electrons increase with increasing temperature due to an increase in the electron-phonon and electron-electron interactions. The increased scattering rate in turn makes the Drude broadening $\Gamma_D$ larger. As a result $\varepsilon_2$, which is proportional to $\Gamma_D$ at longer wavelengths ($\varepsilon_2 \approx \frac{\omega_p^2 \Gamma_D}{\omega^3}$), increases with increasing temperature.

Unlike the imaginary part, the real part of the dielectric permittivity $\varepsilon_1$ (Figure 1a,c,e) changes only marginally with increasing temperature and the trend in $\varepsilon_1$ is noticeable at longer wavelengths (insets of Figure 1a,c,e). Initially, as the temperature is increased up to 200 $^0$C, the real part becomes larger in magnitude. As the temperature is raised further, the real part becomes smaller in magnitude making the film less plasmonic. This behavior is due to two counteracting mechanisms, namely decreasing both the carrier density and electron effective mass, which will be discussed later in the paper. Figure S2a,b shows the room temperature data after each heating cycle. The first cycle improves the film quality by reducing both $\varepsilon_1$ and $\varepsilon_2$. But the subsequent cycles lead to the degradation of the sample properties. This is reflected in the increase in $\varepsilon_2$ as shown in Figure S2b.

Figure 2 shows results on the 200-nm-thick SC film. The imaginary part of the dielectric permittivity $\varepsilon_2$ increases monotonically with the increasing temperature. At longer wavelengths ($\lambda > 900nm$), similar to the thick PC film, we observed a nearly two fold increase in the imaginary part upon heating the sample to 500 $^0$C. This trend can be attributed to the increased scattering rate picture as described above. Similar to the PC films $\varepsilon_1$ only changes marginally with temperature but it becomes slightly larger in magnitude with the increasing temperature (unlike the PC films where $\varepsilon_1$ reduces and increases depending on the temperature range). Figure S2c,d shows the room temperature data of the 200-nm-thick SC film after multiple heating cooling cycles. The imaginary part (Figure S2d) increases after each cycle, reducing the film quality.

Subsequently, we measured the dielectric function of 50-nm-thick film at elevated temperatures. In thinner films, increased surface scattering becomes a dominant contributor to losses. As a result, thinner films have significantly higher losses and hence larger $\varepsilon_2$ than the thicker films. At elevated temperatures, a fundamentally different behavior in the optical constants is observed as shown in Figure 3. In particular, the imaginary part can increase and decrease with increasing temperature (shown in Figure S3a at a wavelength of 1900 nm) unlike for the thicker films where it shows monotonic behavior. Based on the imaginary part behavior, the temperature range is divided into three regions room temperature-200 $^0$C (Figure 3a,d), 200 $^0$C-350 $^0$C (Figure 3b,e) and from 350 $^0$C-500 $^0$C (Figure 3c,f). For the sake of clarity only the data from 1800 nm - 2000 nm is shown in these plots. The DCP terms for these films are shown in Table S6, which can be used to extract the optical constants in the whole spectral range from 370 nm- 2000 nm. Initially, as the temperature is increased from room temperature to 200 $^0$C $\varepsilon_2$ increases as plotted in Figure 3d. But when the temperature is increased from 200

$^0$C to 350 $^0$C $\varepsilon_2$ reduces unlike for the thicker films. For temperatures over 350 $^0$C $\varepsilon_2$ increases again and the samples become extremely lossy (Figure 3f).

The reason for this behavior is as follows: as the temperature is increased from room temperature to 200 $^0$C the scattering rates increase primarily due to an increase in the number of phonons. When the temperature is increased further up to 350 $^0$C, the grain boundaries start to move and the grains merge together forming larger ones, which increases the mean free path and hence reduces the $\varepsilon_2$ [27]. When the temperature is increased over 350 $^0$C the imaginary part again increases and the losses in the samples become extremely large. Note that the y-axis scale in Figure 3f is much larger than that in Figure 3d and Figure 3e. The room temperature measurements performed on the same sample after heat treatment revealed that the optical properties degraded significantly and permanently. This is reflected in the substantial increase in imaginary part as shown in Figure 4a,b.

A similar trend is observed for the 30-nm-thick film. The results at longer wavelengths (1800 nm-2000 nm) are plotted in Figure 5 and also show higher losses compared to the thicker samples (DCP terms for the 30 nm thick film are in Table S7). However, the temperature ranges over which we observe an increase and decrease in $\varepsilon_2$ are different in comparison with the 50-nm-thick film. Based on this trend we divided the experimental temperature range into three regions: room temperature-200 $^0$C (Figure 5a,d), 200 $^0$C-250 $^0$C (Figure 5b,e) and 250 $^0$C-450 $^0$C (Figure 5c,f). As the temperature is raised from room temperature to 200 $^0$C $\varepsilon_2$ increases as shown in Figure 5d. However, when the temperature is increased to 250 $^0$C, $\varepsilon_2$ reduces sharply (Figure 5e) and remains nearly the same at 300 $^0$C (in case of 50-nm-thick film $\varepsilon_2$ continues to reduce until 350 $^0$C). Increasing the temperature further increases $\varepsilon_2$ making it extremely large and at 450 $^0$C it increases by nearly a factor of four compared to the room temperature data (Figure 5f). Figure S3b shows this behavior in $\varepsilon_2$ at a wavelength of 1900 nm. Similar to the above, this is due to the two counteracting mechanisms of increasing electron-phonon interactions and the grain boundary movements, which increase the mean free path. Figure 4c and 4d show the room temperature optical constants of the 30-nm-thick sample both before and after the heat treatment. A substantial increase in $\varepsilon_2$ is seen after the heat treatment revealing that the optical properties have degraded similar to that seen in 50-nm-thick film.

The AFM images (Figures S4 &S5) of thin films before and after heating revealed that the surface roughness of the films increased after the heat treatment. The increased roughness increases the surface scattering, thus leading to the observed increase in the imaginary part. Further, the optical images of these films (Figure S6) showed that several cracks were formed in the film after the heat treatment. Despite these cracks, we fit the obtained VASE data assuming the films to be continuous. Good fits were obtained with a MSE <1.5 (Tables S6 & S7) suggesting that the assumption made on film continuity is reasonable. Although thin film fitting is accurate at these high temperatures, it should be noted that the film morphology starts changing. Consequently, modeling of the films at even higher temperatures requires further modification[28, 29].

Using the temperature-dependent optical constants we have estimated the performance of gold-based plasmonic systems at elevated temperatures. Specifically, we have estimated the propagation lengths of surface plasmon polaritons (SPP) at the air-gold interface[5], and the quality factor of localized surface plasmon resonance ($Q_{LSPR}$). The SPP propagation length is defined as the distance over which the intensity of the SPP decays by a factor of *e*, and $Q_{LSPR}$ is defined as the ratio of enhanced local field to the incident field[30]. For a spherical particle in the quasistatic regime, it can be shown that $Q_{LSPR} = \frac{-\varepsilon_1}{\varepsilon_2}$ [30-32]. The computed results of SPP propagation lengths and $Q_{LSPR}$ for a spherical particle obtained using the data from 200-nm-thick films are shown in Figure S7. At 500 $^0$C a significant reduction in the both the propagation lengths and the $Q_{LSPR}$ by over 47% and 40% is observed in PC and SC films, respectively. Thus, SC films show more thermal stability compared to the PC films. For thinner films, we see an even stronger reduction in the propagation lengths and $Q_{LSPR}$ by $\sim$ 50% and 70% in 50-nm and 30-nm-thick samples, respectively. Table 1 shows the propagation lengths of SPPs and $Q_{LSPR}$ of the 200-nm-thick PC and SC films at 820 nm wavelength. Table 2 shows the same for thinner films at the same wavelength. Note that the propagation lengths and $Q_{LSPR}$ at room temperature in the thinner films are significantly smaller than that in thicker samples and the observed relative change is also larger.

## THEORY

In this section we compare the experimental results fitted using the DCP model with the theoretical predictions. At longer wavelengths, where the inter-band transitions become insignificant, the observed temperature dependencies in the optical constants are due to the two Drude terms: Plasma frequency $\omega_p$ and the Drude damping $\Gamma_D$. The temperature dependencies of these two terms are primarily due to the following factors: 1) the decrease in the carrier density due to volume expansion, 2) the decrease of the effective mass of the free electrons in the metal and 3) the increase in the electron-phonon interaction with increasing temperature. The plasma frequency $\omega_p$ is dependent on the carrier density ($N$) and the effective mass ($m^*$) of the electrons according to the relation:

$$\omega_p^2 = \frac{Ne^2}{m^* \varepsilon_0} \ . \qquad (1.4)$$

The carrier density $N$ reduces with increase in temperature due to volume thermal expansion according to

$$N = \frac{N_0}{1 + \gamma(T - T_0)} \ , \qquad (1.5)$$

where $\gamma$ is the volume thermal expansion coefficient. On the other hand the effective mass $m^*$ in metals has been reported to decrease with increasing temperature[33]. The decrease in $m^*$ increases the plasma frequency whereas the decrease in $N$ counteracts it. The interplay between these two mechanisms dictates the observed behavior in the plasma frequency. The temperature dependence of $\omega_p$ for 200-nm-thick PC (SC) film is shown in Figure 6a (S8a). The error bars in the plots show the 90% confidence limits obtained from the fits. Based on these experimental findings we conclude that the decrease in $m^*$ (increase in the plasma frequency) is the dominant mechanism compared to the change in $N$ for temperatures below 200 $^0$C and for even higher temperature the decrease in $N$ (reducing the plasma frequency) is the dominant mechanism. A similar increase in the plasma frequency up to 200 $^{0}$C is observed in the thinner films (Table S6 and Table S7). On the other hand, for the 200-nm-thick SC film the plasma frequency monotonically increases indicating that the reducing effective mass is the dominant mechanism throughout the measured temperature range (Figure S8a). A similar increase in the plasma frequency was reported in silver films[17].

The Drude damping term $\Gamma_D = \frac{\hbar}{\tau_D}$, where $\hbar$ is the reduced Planck constant and $\tau_D$ is the electron relaxation time, depends on the electron-electron ($\Gamma_{ee}$) and electron-phonon ($\Gamma_{e\phi}$) scattering mechanisms:

$$\Gamma_D = \Gamma_{ee} + \Gamma_{e\phi} \ , \tag{1.6}$$

or

$$\frac{1}{\tau_D} = \frac{1}{\tau_{ee}} + \frac{1}{\tau_{e\phi}} \ . \tag{1.7}$$

Where $\frac{1}{\tau_{ee}}$ and $\frac{1}{\tau_{e\phi}}$ are given by[34-39]

$$\frac{1}{\tau_{ee}} = \frac{1}{12}\pi^3\Gamma\Delta\left(\frac{1}{\hbar E_F}\right)\left[(K_BT)^2 + \left(\frac{\hbar\omega}{2\pi}\right)^2\right] , \tag{1.8}$$

$$\frac{1}{\tau_{e\phi}} = \frac{1}{\tau_o}\left[\frac{2}{5} + 4\left(\frac{T}{\theta}\right)^5 \int_0^{\frac{\theta}{T}} \frac{z^4}{e^z - 1} dz\right] . \tag{1.9}$$

Here $\Gamma$, $\Delta$, $E_F$, $\theta$ and $\tau_o$ are the average scattering probability over the Fermi surface, the fractional Umklapp Scattering, the Fermi energy of free electrons, the Debye temperature and a material dependent constant, respectively.

The Debye temperature for gold is 170 K. So $\frac{\theta}{T}$, the upper limit of the integral in Eqn (1.7), is $< 1$ for the whole temperature range studied in this work. Hence the exponential in the denominator of the integrand can be approximated using the

Taylor's series as $e^{z} \approx 1 + z$. This reduces the expression of electron-phonon scattering to

$$\frac{1}{\tau_{e\phi}} = \frac{1}{\tau_0}\left[\frac{2}{5} + \frac{T}{\theta}\right]. \tag{1.10}$$

Although the electron-electron scattering has a quadratic dependence on the temperature, its contribution due to the temperature dependent term is weak compared to the frequency dependent term for the optical frequencies. However, a frequency independent $\Gamma_D$ (hence a frequency independent scattering rate) is widely used in modelling the optical constants of metal films. We therefore treat the contribution due to electron-electron scattering as independent of both temperature and frequency and obtain good fits for the Drude broadening. These results for the 200-nm-thick PC (SC) film are shown in Figure 6b (S8b). Qualitatively, these temperature dependences of the plasma frequency and the Drude broadening can be understood from the Drude equation

$$\hat{\varepsilon}(\omega, T) = \varepsilon_{\infty} - \frac{\omega_p(T)^2}{\omega^2 + i\Gamma_D(T)\omega} \ . \tag{1.11}$$

By differentiating the real and imaginary parts of the dielectric function with respect to temperature in Eqn (1.11) (and assuming no temperature dependence of $\varepsilon_{\infty}$ ) we get

$$\frac{d\omega_p}{dT} = \omega^2 \frac{\left(\frac{\Gamma_D^2}{\omega^2} - 1\right)\frac{\partial \varepsilon_1}{\partial T} + 2\frac{\Gamma_D}{\omega}\frac{\partial \varepsilon_2}{\partial T}}{2\omega_p} \tag{1.12}$$

and

$$\frac{d\Gamma_D}{dT} = \omega^3 \frac{1 + \Gamma_D^2/\omega^2}{\omega_p^2}\left(\frac{\Gamma_D}{\omega}\frac{\partial \varepsilon_1}{\partial T} + \frac{\partial \varepsilon_2}{\partial T}\right) \tag{1.13}$$

As $\Gamma_D \ll \omega$, $\Gamma_D/\omega \to 0$ and Eqns (1.12) and (1.13) can be approximated as

$$\frac{d\omega_p}{dT} \approx -\frac{\omega^2}{2\omega_p}\frac{\partial \varepsilon_1}{\partial T} \tag{1.14}$$

$$\frac{d\Gamma_D}{dT} \approx \frac{\omega^3}{\omega_p^2}\frac{\partial \varepsilon_2}{\partial T} \tag{1.15}$$

Therefore, the temperature dependence of plasma frequency is proportional to the negative of the temperature derivative of $\varepsilon_1$, while that of $\Gamma_D$ is proportional to the temperature derivative of $\varepsilon_2$ . Comparing the observed temperature dependence in $\omega_p$ and $\Gamma_D$ (shown in Figure 6 for the 200-nm-thick PC films) with the plots shown in Figure 1 (e and f) indeed confirms these relations.

The oscillator strengths and the oscillator damping coefficients increase monotonically with increasing temperatures. These temperature dependencies can be described using Bose–Einstein phenomenological models. For oscillator strengths, we use the following form to describe its temperature dependence[40,41]:

$$C(T) = C_0 \coth\left(\frac{\theta}{2T}\right) + \alpha \tag{1.16}$$

Where $C_0$ and $\alpha$ are the material dependent parameters and are supplied as fit parameters. We use the same form of the empirical expression to describe the gold oscillator strengths.

Similar to the oscillator strengths, the oscillator damping can be described using the phenomenological model[42]:

$$\gamma(T) = \gamma_0 \coth\left(\frac{\theta}{2T}\right) + \gamma_1 \tag{1.17}$$

Here $\gamma_0$ and $\gamma_1$ are supplied as the fit parameters. Good fits for the experimental data were obtained using these phenomenological models for the oscillator strengths and oscillator dampings. The results for 200 nm thick PC film using these models are shown in Figure 7a-d.

For the case of oscillator energies, one of the oscillator energies decrease with increasing temperature while the other one increases with increasing temperature. A similar trend in the oscillator energies in gold has been reported in an earlier paper[14,43]. Figures 7e, 7f show these results for the 200 nm thick PC film.

## CONCLUSIONS

To conclude, we have measured the optical properties of thin gold films at elevated temperatures and provided analytical models to describe the temperature-dependent dielectric function. Our findings show that the imaginary part of the dielectric function changes significantly with the increasing temperature, while the real part remains almost intact. The observed increase in the imaginary part (by nearly 3-4 times in thinner films and 2 times in thicker samples) significantly reduces the propagation length of the surface plasmon polaritons at the gold-air interface and the quality factor of the localized surface plasmon resonances in spherical particles. For thin gold films we observed both a decrease and increase of the imaginary part depending on the temperature range. Furthermore, the thin gold films were permanently damaged when the temperature was increased to 500 $^0$C. We utilized experiment-fitted models to describe the temperature-dependent gold dielectric function as a sum of one Drude and two Critical Point terms (parameters shown in Tables S1-S7). Our experimental results indicate that temperature-dependent deviations in the gold optical constants are quite significant. This is an important finding for local-heat assisted applications such as nanolasers and spasers, plasmonic sensors, photothermal therapy with nanoparticles, photocatalysis and heat-assisted magnetic recording. Henceforth, the causal analytical models developed in this work shall be of critical assistance for accurate multiphysics modelling of gold-based high temperature nanophotonic and plasmonic devices operating in steady-state or dynamic regimes.

## METHODS

### Sample preparation

Polycrystalline samples with various thicknesses were deposited on a 1-mm thick glass substrate using an electron beam evaporator (CHA Industries Model 600) at room temperature. The pressure during the deposition process was of the order of $\sim 1\times10^{-6}$ torr. No adhesion layer was used during the deposition. The 200-nm-thick SC films were purchased from Phasis Sarl. These samples were deposited on Mica at 400 $^0$C in order to obtain the crystalline phase.

### Experimental setup

Details of the experimental setup are provided in the supporting information

# AUTHOR INFORMATION

### Corresponding authors

*E-mail: uguler@nanometatech.com

*E-mail: shalaev@purdue.edu

[#] H.R and U.G contributed equally to the work.

## Notes

The authors declare no competing financial interest.

# ACKNOWLEDGEMENTS

The authors would like to thank Prof. Ali Shakouri, Amirkoushyar Ziabari, Yeerui Koh and Amr Mohammed for their help with the experimental setup. This work was supported in part by NSF MRSEC (DMR-1120923) and NSF SBIR (IIP-1416232).

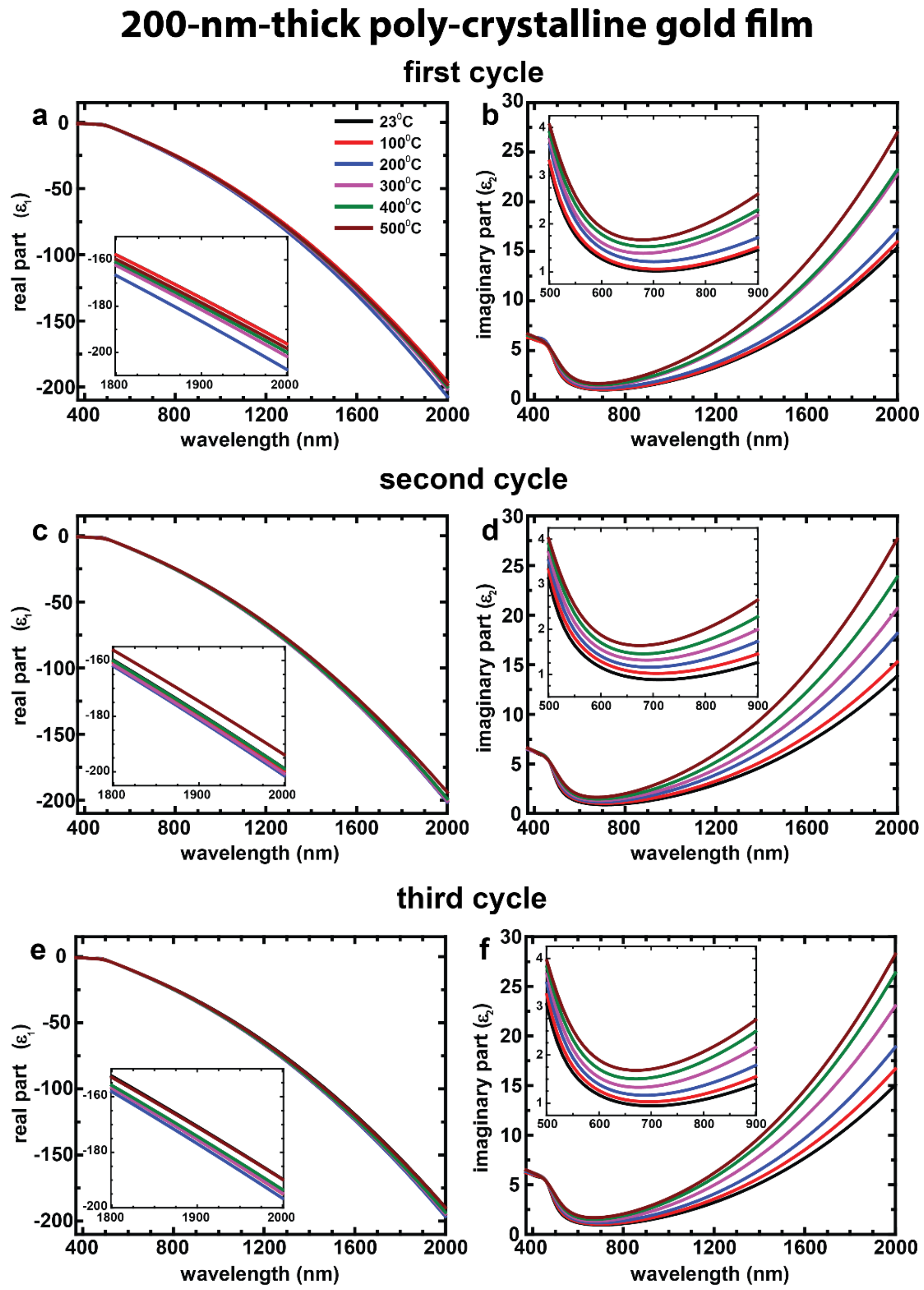

**Figure 1:** Optical constants of 200 nm thick PC films. (a) and (b) show the real and imaginary parts of the dielectric function, respectively for the first cycle. Likewise (c), (d) and (e), (f) are those of second and third cycle. Different colors represent the dielectric functions at different temperatures (legend in (a) shows the color coding). The imaginary part increases monotonically with increasing temperature whereas the real part decreases with increasing temperature up to 200 $^0$C and increases when the temperature is increased further. The same trend is observed for all three cycles. Insets show the real and imaginary parts for a selected wavelength range.

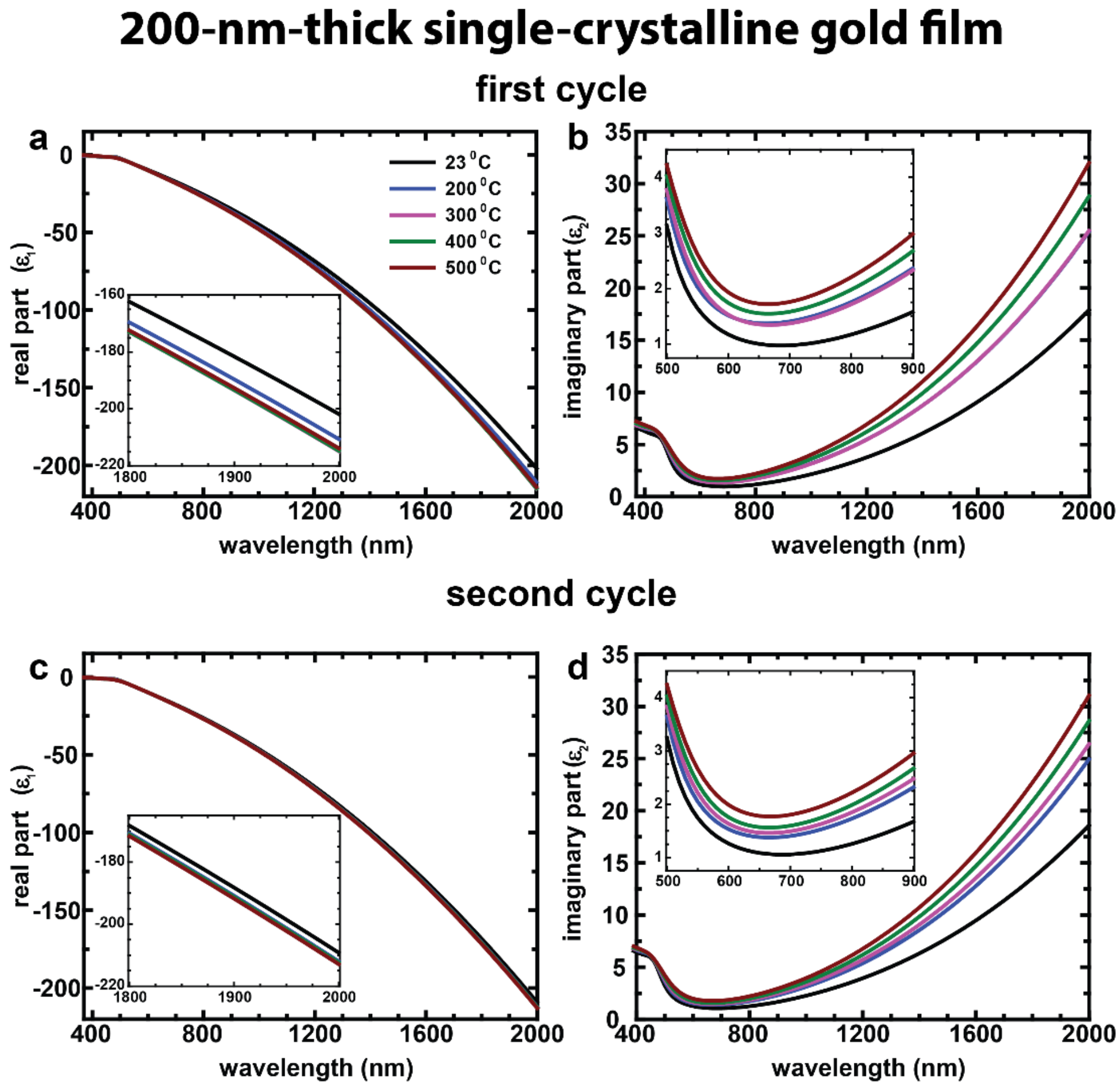

**Figure 2:** Optical constants of 200 nm thick SC films. (a) and (b) show the real and imaginary parts of the dielectric function, respectively for the first cycle. Likewise (c), (d) are those of the second cycle. Different colors represent the dielectric functions at different temperatures (legend in (a) shows the color coding). Similar to the PC films the imaginary part increases monotonically with increasing temperature whereas the real part decreases with increasing temperature and saturates at 500 $^0$C. The same trend is observed for both cycles. Insets show the real and imaginary parts for a selected wavelength range.

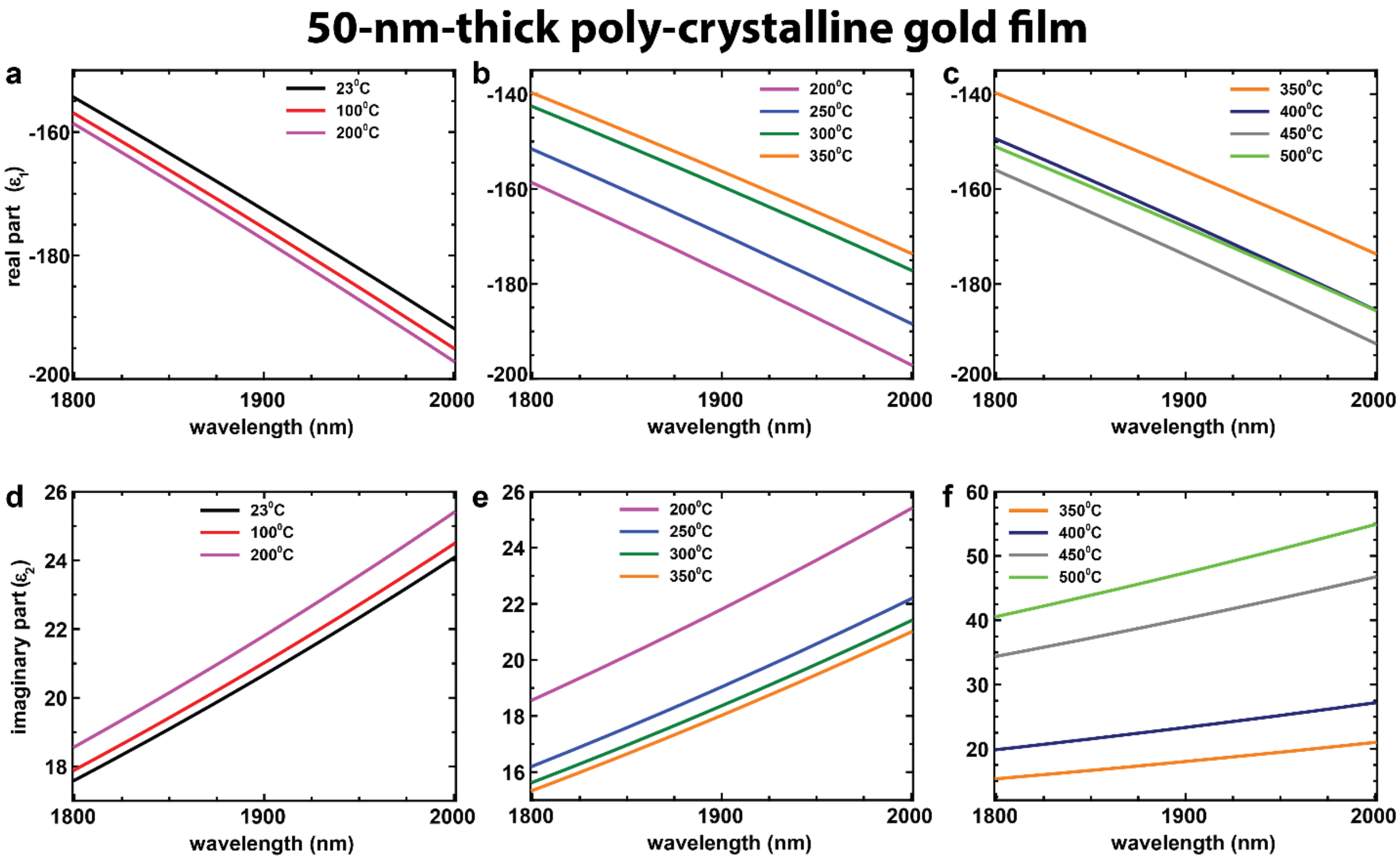

**Figure 3:** Optical constants of 50 nm thick poly crystalline films. (a), (b) and (c) show the real part and (d), (e) and (f) show the imaginary part of the dielectric function for different temperature regions. Different colors correspond to dielectric functions at different temperatures (shown in the legend of each figure). As the temperature is increased from room temperature the imaginary part (d) increases up to 200 $^0$C. But for the temperature range from 200 $^0$C- 350 $^0$C the imaginary part (e) reduces, unlike the thicker films. Increasing the temperature further increases the imaginary part drastically reducing the film quality significantly as shown in (f). The real part also displays increasing and deceasing behavior with temperature, depending on the temperature range.

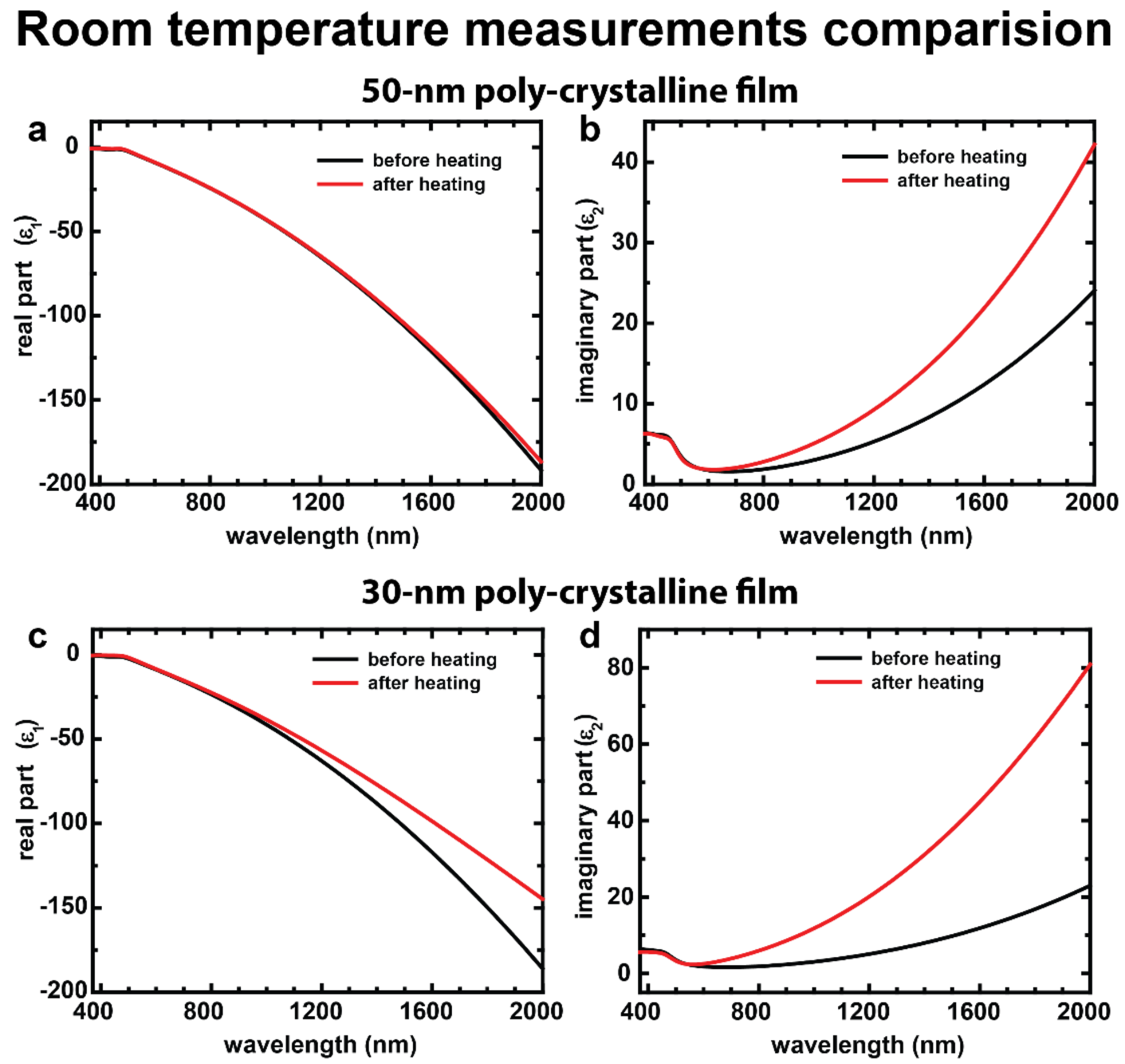

**Figure 4**: Room temperature measurements on the 50 nm and 30 nm thick gold film. The black and red curves represent the room temperature dielectric function on the same sample before and after heating, respectively. Both the real part (a,c) and the imaginary part (b,d) increase after heating the sample.

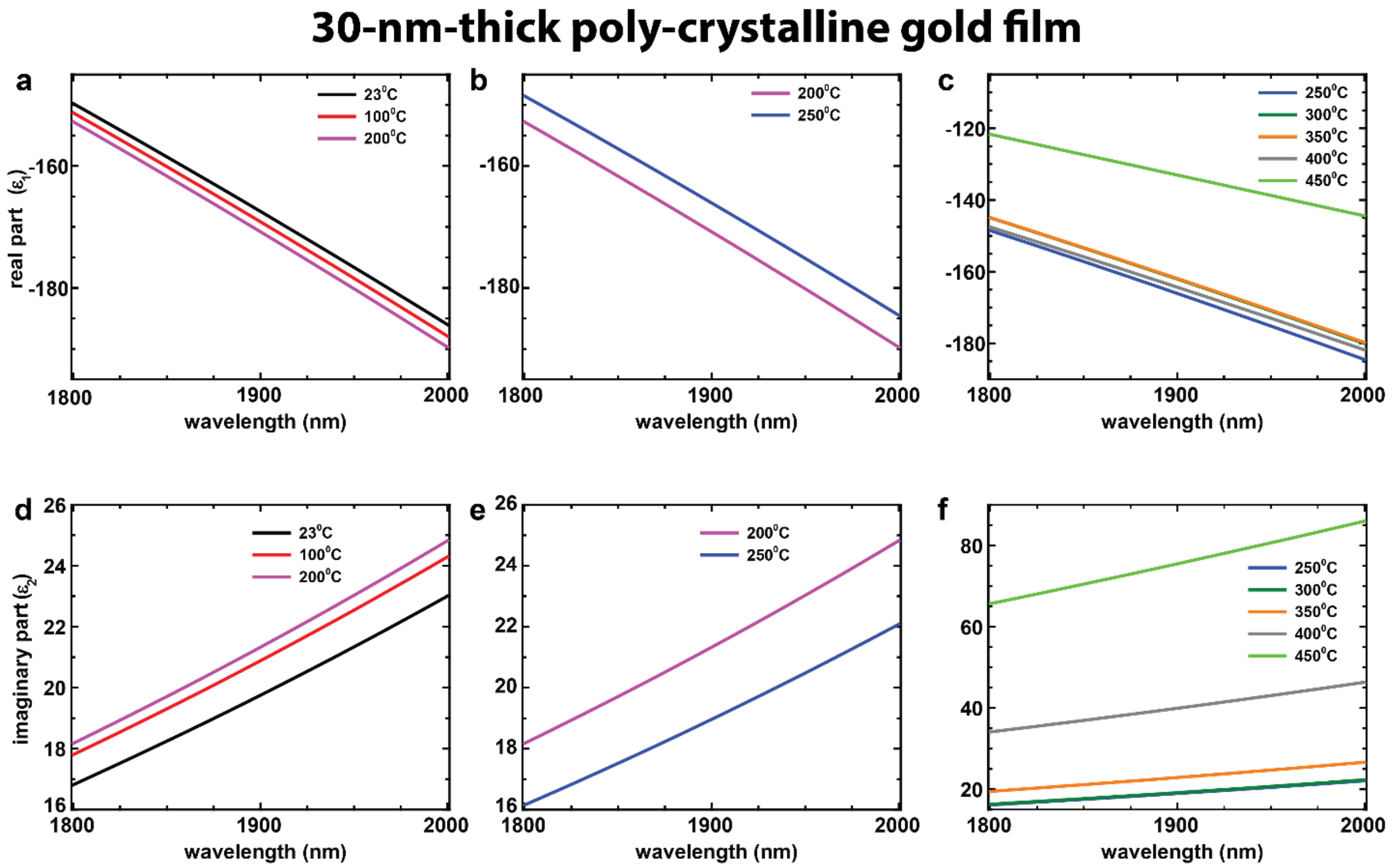

**Figure 5:** Optical constants of 30 nm thick poly crystalline films. (a), (b) and (c) show the real part and (d), (e) and (f) show the imaginary part of the dielectric function for different temperature regions. Different colors correspond to dielectric functions at different temperatures (shown in the legend of each figure). Initially, the imaginary part (d) increases as the temperature is increased from room temperature to 200 $^0$C. Similar to the 50 nm thick samples, the imaginary part (e) reduces when the temperature is increased to 250 $^0$C. When the temperature is increased over 300 $^0$C the imaginary part (f) increases and becomes extremely large. The real part also displays increasing and decreasing behavior with temperature, depending on the temperature range.

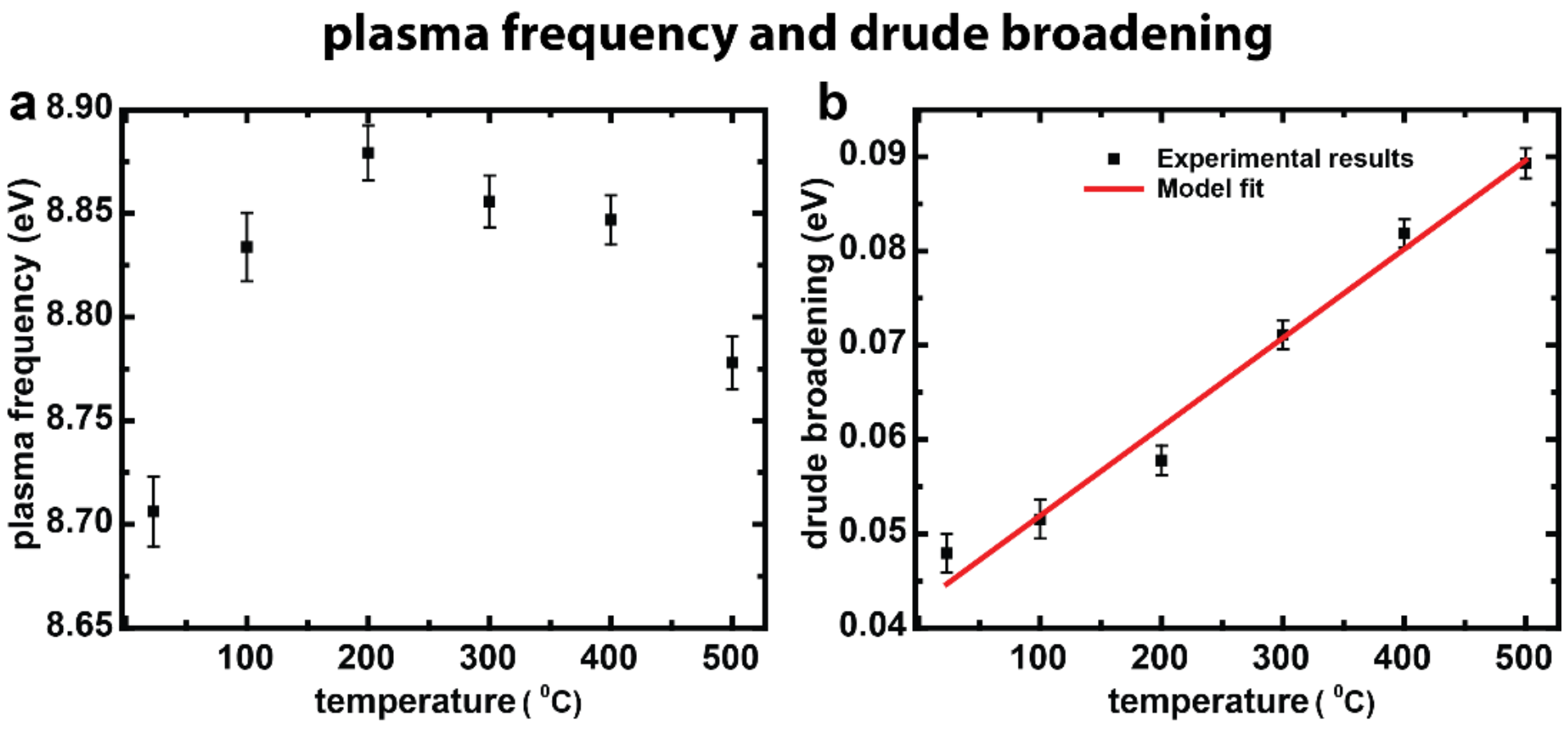

**Figure 6:** Plasma frequency (a) and Drude broadening (b) of 200 nm PC films. Depending on the temperature range the plasma frequency either increases or decreases. On the other hand, the Drude broadening increases monotonically with increasing in temperature. The red curve is the fit obtained using Eqn (1.10).

**critical point terms**

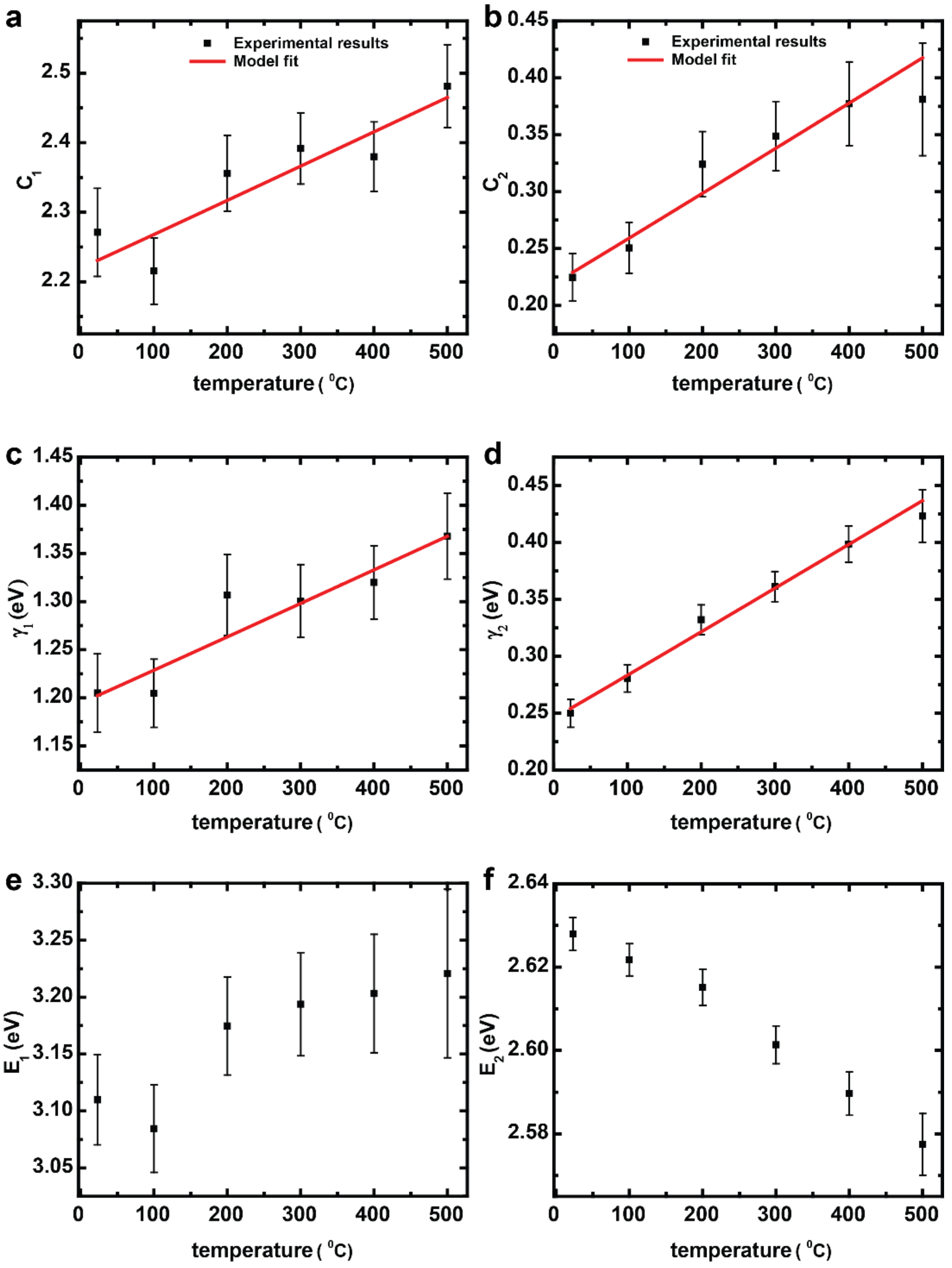

**Figure 7:** Oscillator strengths (a,b), Oscillator dampings (c,d) and Oscillator energies (e,f) of 200 nm PC films. The red curve shows the fit obtained using the empirical expressions discussed in the Theory section.

## Table 1

| | SPP Prop Length at 23 $^0$C | SPP Prop Length at 500 $^0$C | % change in Prop Length | $Q_{LSPR}$ ($-\varepsilon_1/\varepsilon_2$) at 23 $^0$C | $Q_{LSPR}$ ($-\varepsilon_1/\varepsilon_2$) at 500 $^0$C | % change in $Q_{LSPR}$ |
|---|---|---|---|---|---|---|
| 200 nm PC Au | 72.94 µm | 38.41 µm | 47.3 % | 22.93 | 11.93 | 48% |
| 200 nm SC Au | 74.86 µm | 44.66 µm | 40.3 % | 21.35 | 12.37 | 42.1 % |

**Table 1:** Comparison of SPP propagation lengths and $Q_{LSPR}$ at Room temperature and 500 $^0$C for 200 nm thick films.

## Table 2

| | SPP Prop Length at 23 $^0$C | SPP Prop Length at 450 $^0$C | % change in Prop Length | $Q_{LSPR}$ ($-\varepsilon_1/\varepsilon_2$) at 23 $^0$C | $Q_{LSPR}$ ($-\varepsilon_1/\varepsilon_2$) at 450 $^0$C | % change in $Q_{LSPR}$ |
|---|---|---|---|---|---|---|
| 50 nm PC Au | 43.08 µm | 25.11 µm<br>#(21.58 µm) | 41.7%<br>#(49.91%) | 13.34 | 7.48<br>#(6.47) | 44%<br>#(49.5%) |
| 30 nm PC Au | 40.49 µm | 11.35 µm | 72 % | 13.05 | 3.50 | 73.2% |

**Table 1:** Comparison of SPP propagation lengths and $Q_{LSPR}$ at Room temperature and 450 $^0$C for 50 nm and 30 nm thick films (# represents computed values of propagation lengths and $Q_{LSPR}$ at 500 $^0$C).

# Supporting Information

## Experimental setup:

To study the temperature dependent optical properties, a heating stage was integrated onto our Variable Angle Spectroscopic Ellipsometer (VASE) setup. However, at temperatures over 400 - 450 $^{0}$C the background thermal radiation becomes strong enough and saturates the detector. In order to reduce the intensity of the thermal radiation reaching the detector a pinhole was introduced in the reflected beam path (shown in Figure S1). This pinhole suppresses most of the background thermal radiation while allowing most of the reflected light more than 85 % of the reflected beam intensity) to pass through.

For all our samples, prior to the temperature dependent measurements a calibration was done using a Si/ $SiO_2$ wafer that was purchased from J.A. Woolam company. As most of the reflected beam reached the detector (> 85 %), the introduction of pinhole didn’t lead to any noticeable difference in the optical properties compared to the case where there was no pinhole.

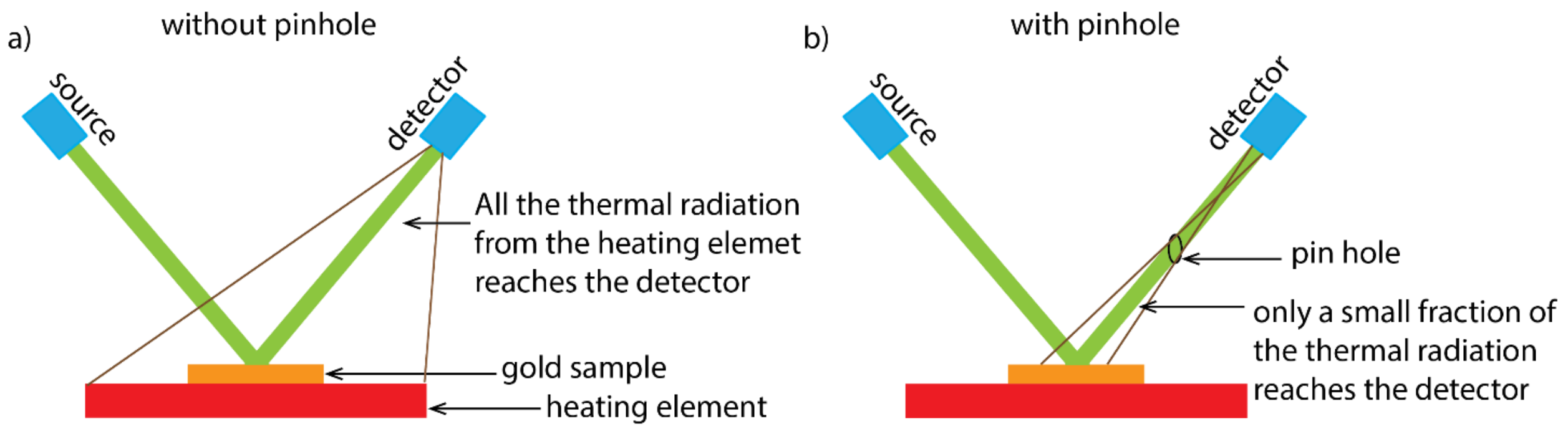

**Figure S1**: Schematic of the experimental setup without (a) and with (b) pinhole in the reflected beam path. Introducing the pinhole (b) significantly suppresses the intensity of background thermal radiation reaching the detector while still allowing most of the reflected beam to pass through.

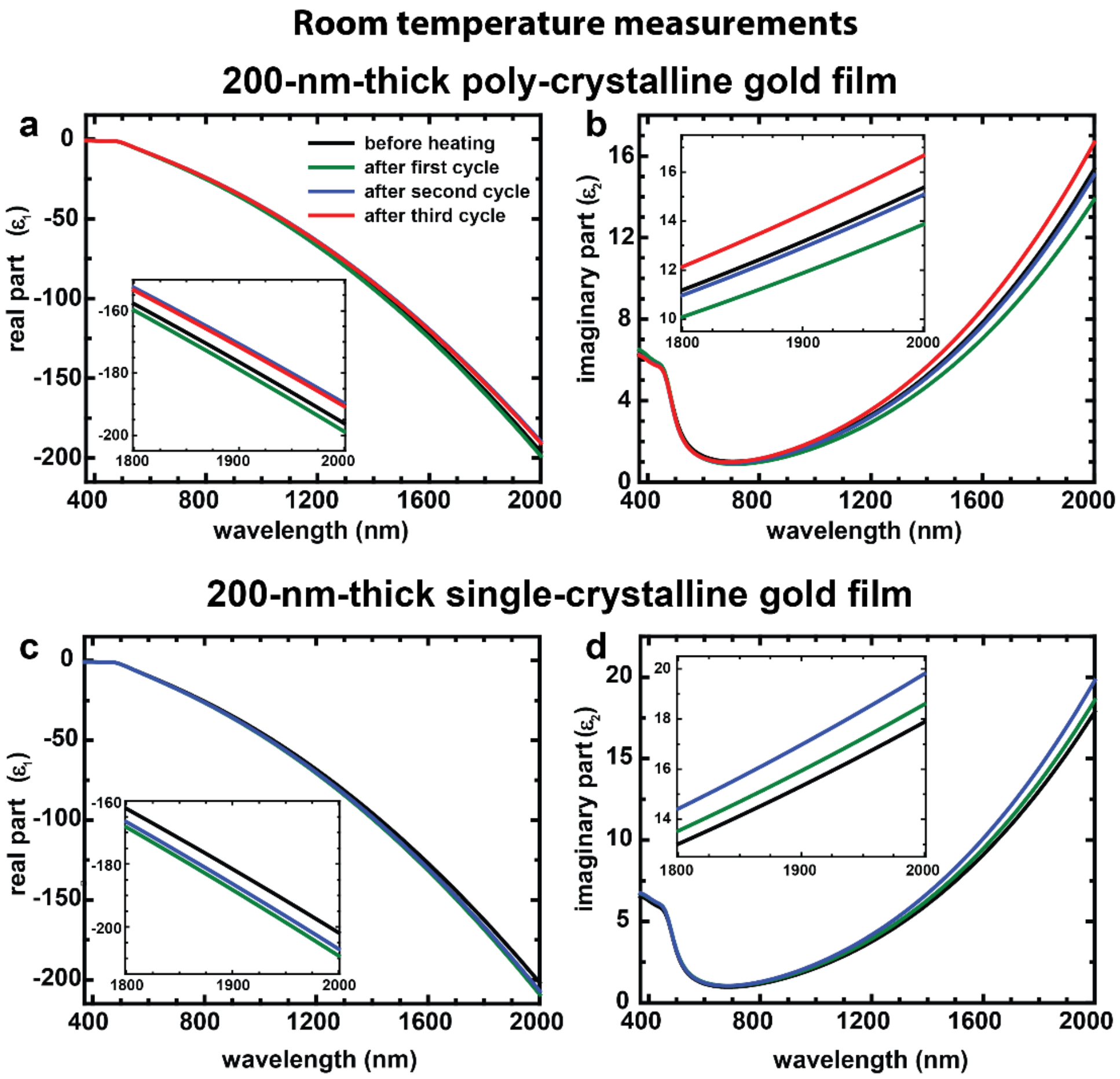

**Figure S2:** Room temperature dielectric function of the 200-nm-thick poly-crystalline (a,b) and single-crystalline (c,d) films after each cycle. After the first heating cycle the imaginary part reduces (green curves in (a) and (b)) thus improving the film quality. But when the film is subjected to subsequent heat cycles the imaginary part start to increase, gradually degrading the film quality (blue and red curves in (b)). For the case of single crystalline films, the imaginary part increases after each cycle (green and blue curves in (d)). In both the samples the real part only changes marginally with repeated heating.

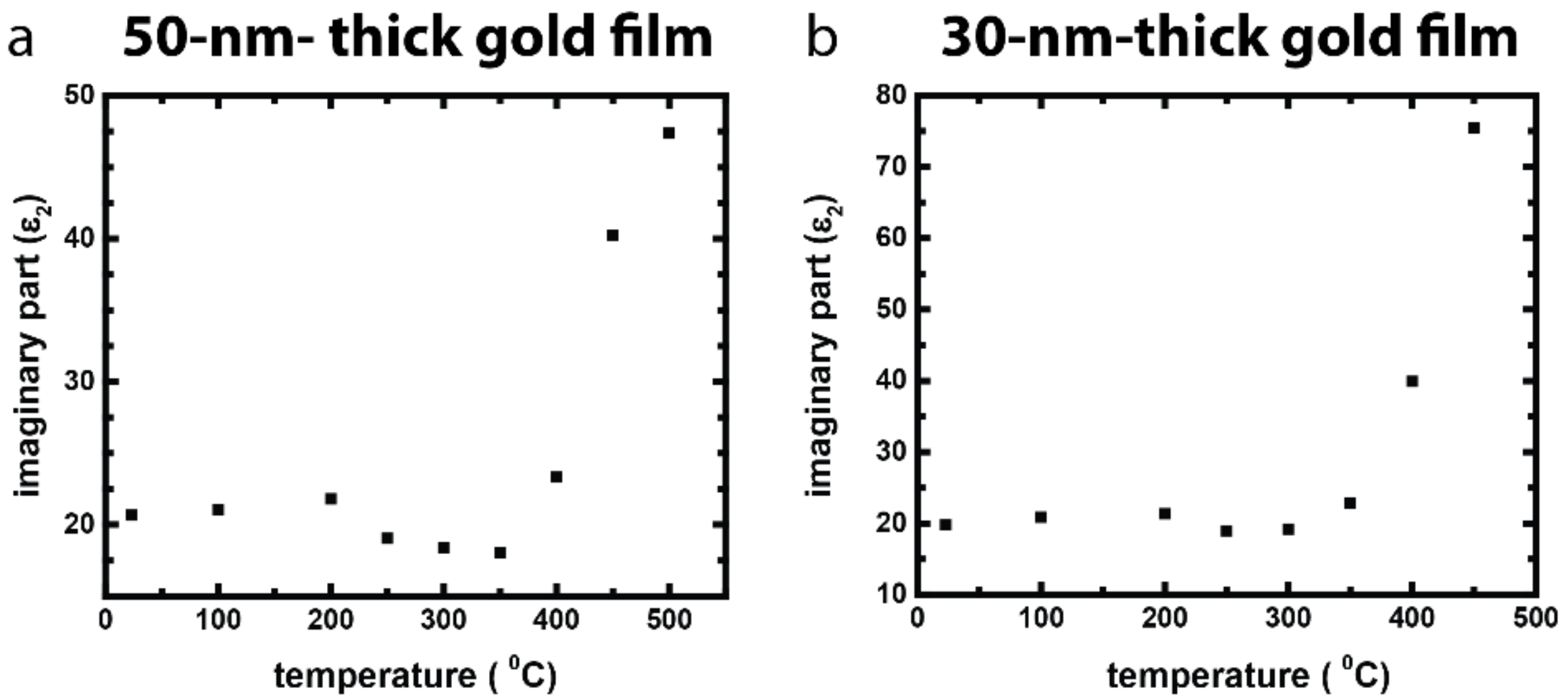

**Figure S3:** Temperature dependence of the imaginary part of the dielectric function at 1900 nm wavelength for 50-nm-thick (a) and 30-nm-thick films (b). Depending on the temperature range the imaginary part either increase or decreases.

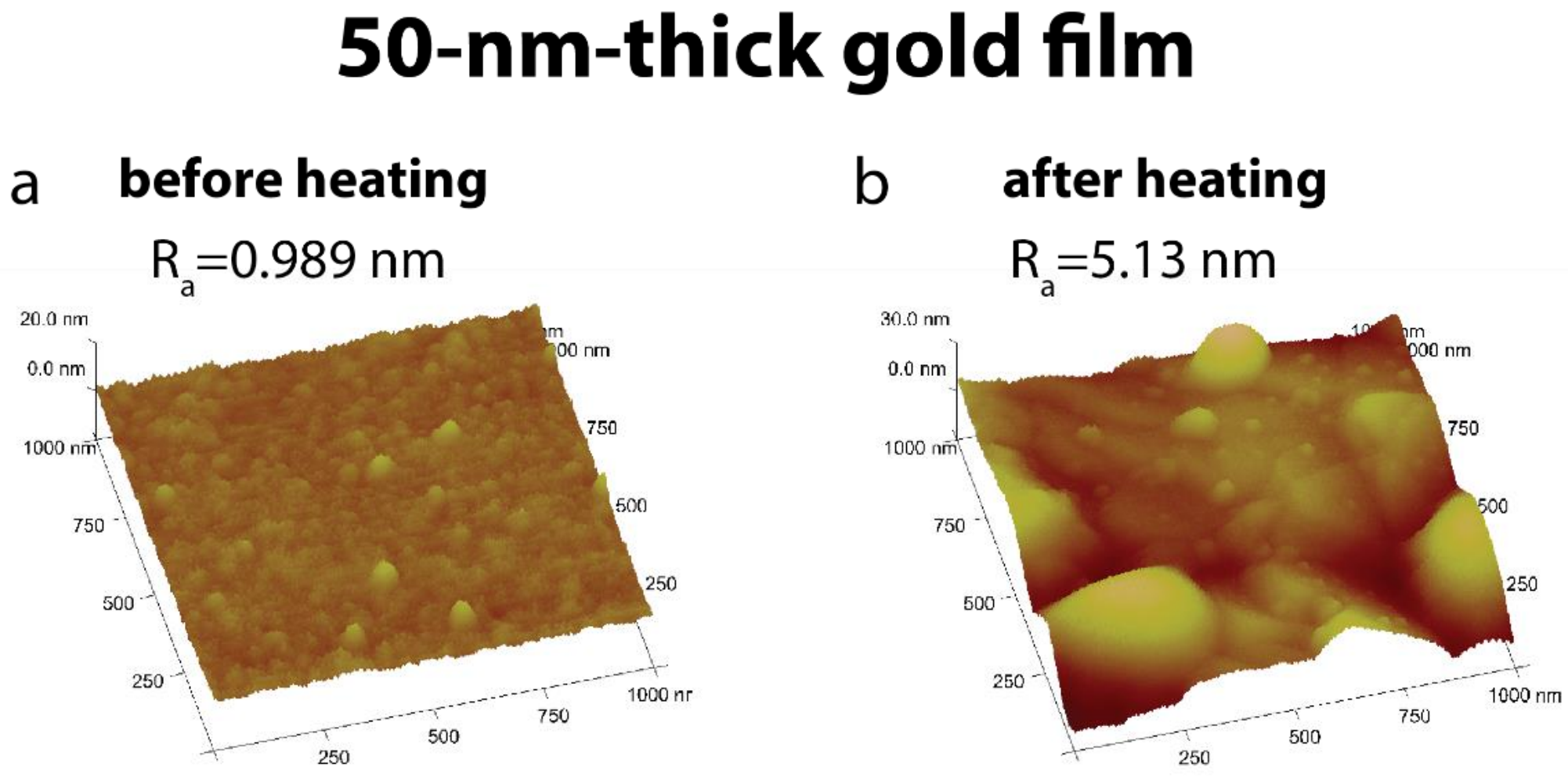

**Figure S4**: AFM images of 50-nm-thick poly-crystalline films. The mean roughness (Ra), which represents the average of the deviations from the center plane, after the heat treatment (b) increased significantly compared to the same samples before heating (a).

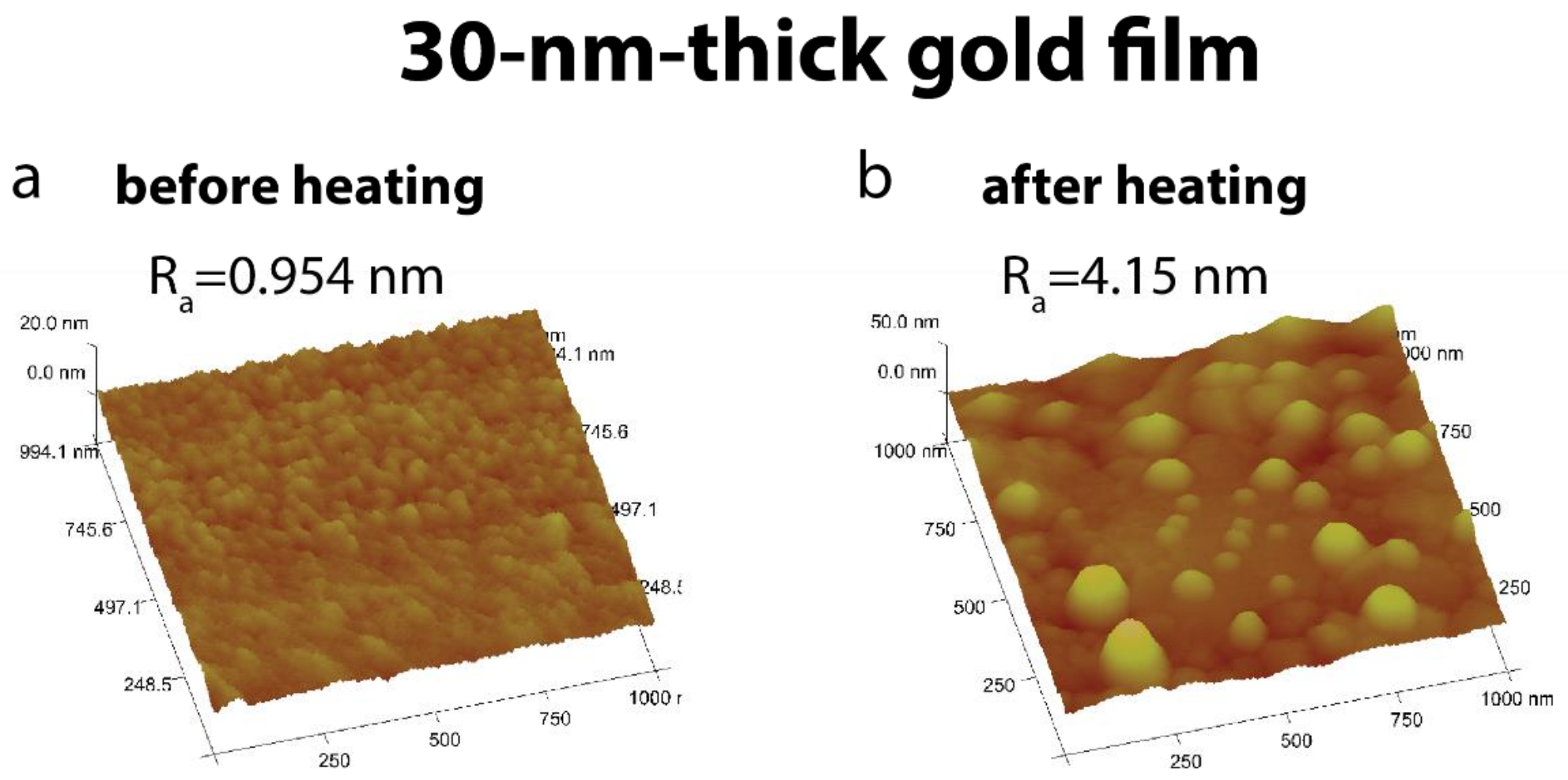

**Figure S5**: AFM images of 30-nm-thick poly-crystalline films. Similar to the 50-nm-thick films the mean roughness ($R_a$) increased after the heat treatment (b) compared to the same samples before heating (a).

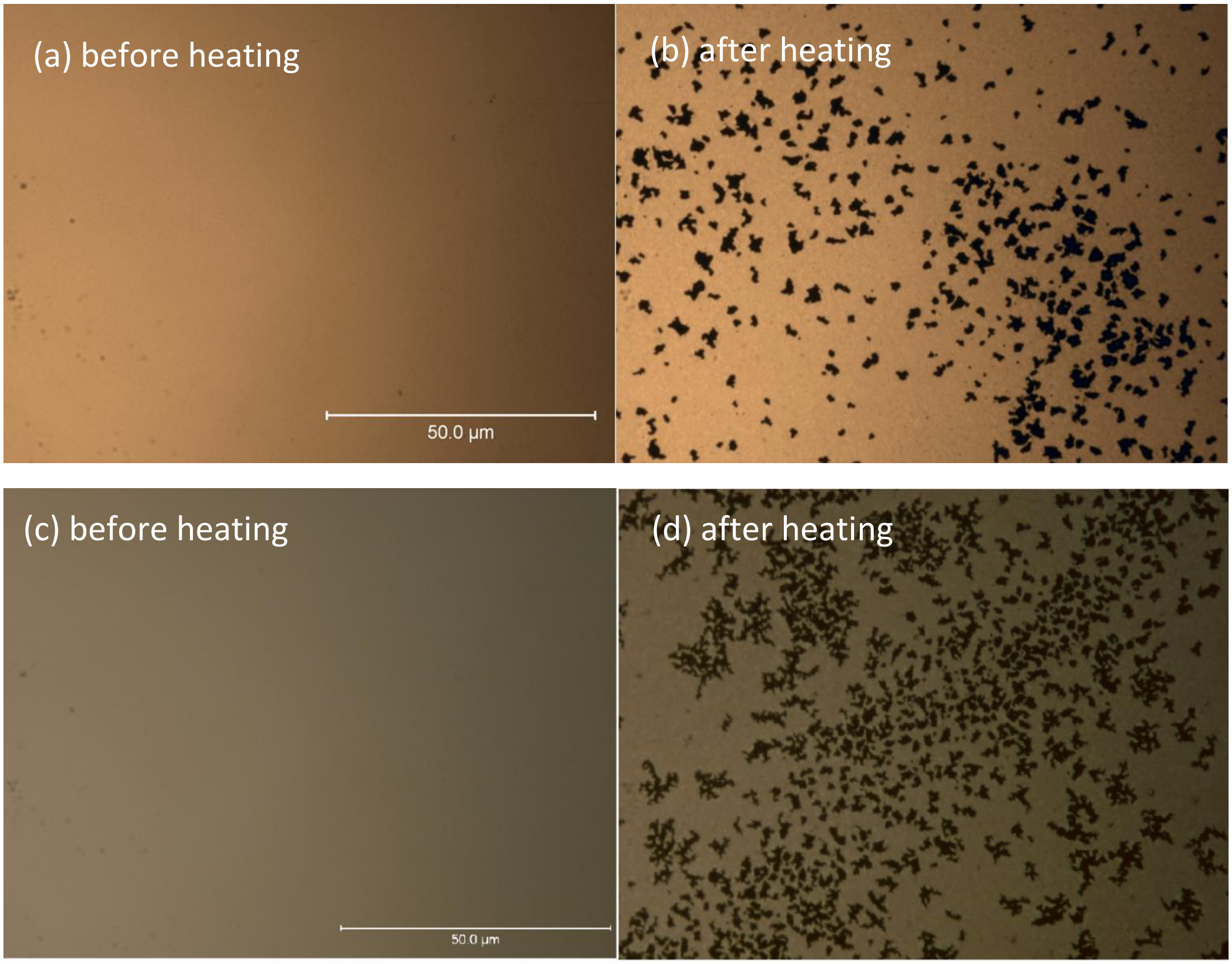

**Figure S6:** Optical images of the 50-nm- and 30-nm-thick films. Images before (a) and after heating (b) confirm that the 50-nm-thick film has degraded significantly. Several cracks can be seen in the film after heating (b). Similar behavior is seen in 30-nm-thick films (c, d).

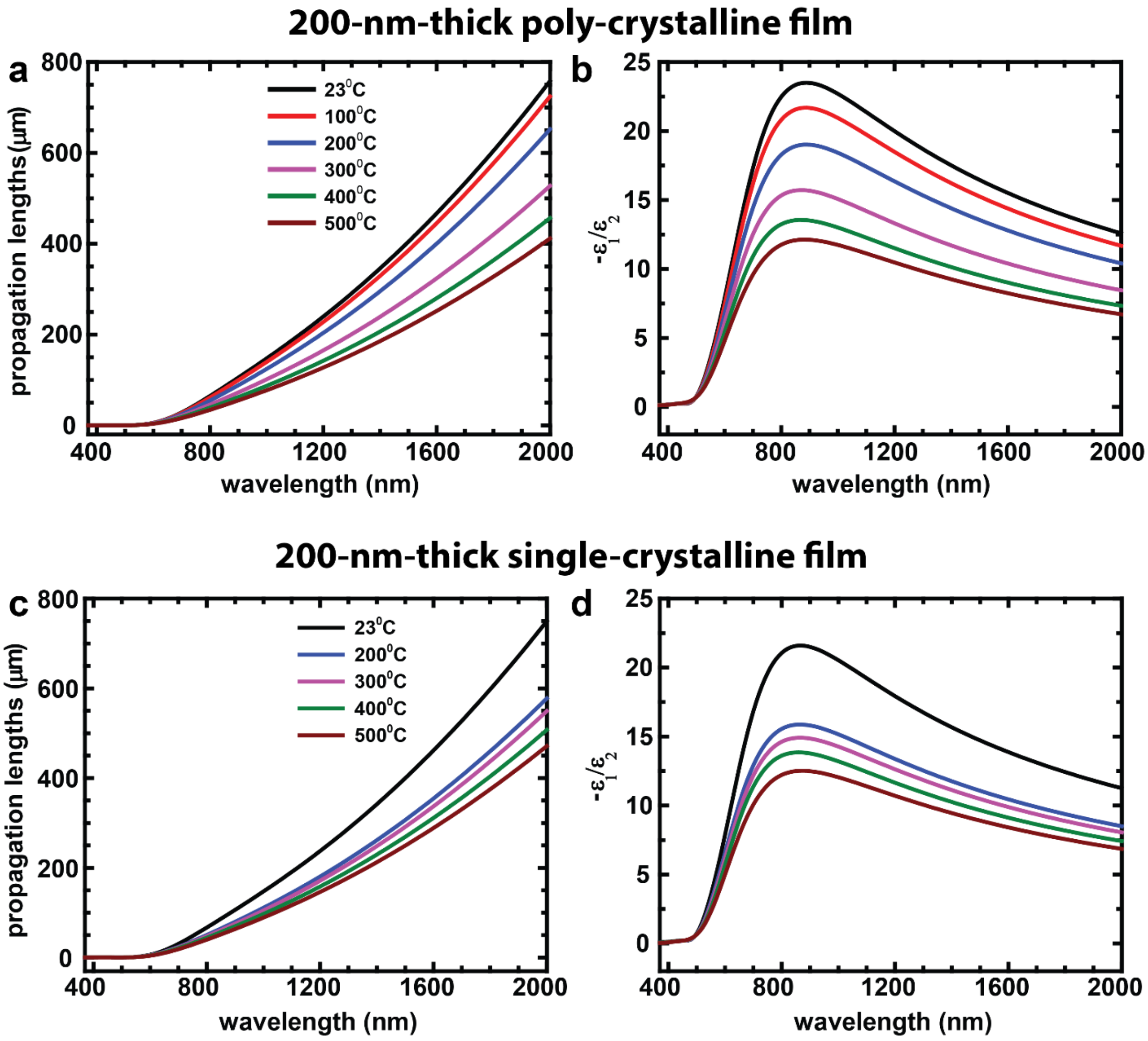

**Figure S7:** Computed values of temperature dependent SPP propagation lengths and $Q_{LSPR}$ $\left(\frac{-\varepsilon_1}{\varepsilon_2}\right)$ using the optical constants of 200-nm-thick poly-crystalline (a,b) and single crystalline (c,d) gold films. Legends in Figures (a) and (c) show the color coding. In both cases, the propagation lengths and $Q_{LSPR}$ reduce by nearly a factor of two compared to the room temperature results when the temperature is raised to 500 $^0$C. These results were computed using the third cycle and second cycle optical constants for the polycrystalline and single crystalline gold films, respectively.

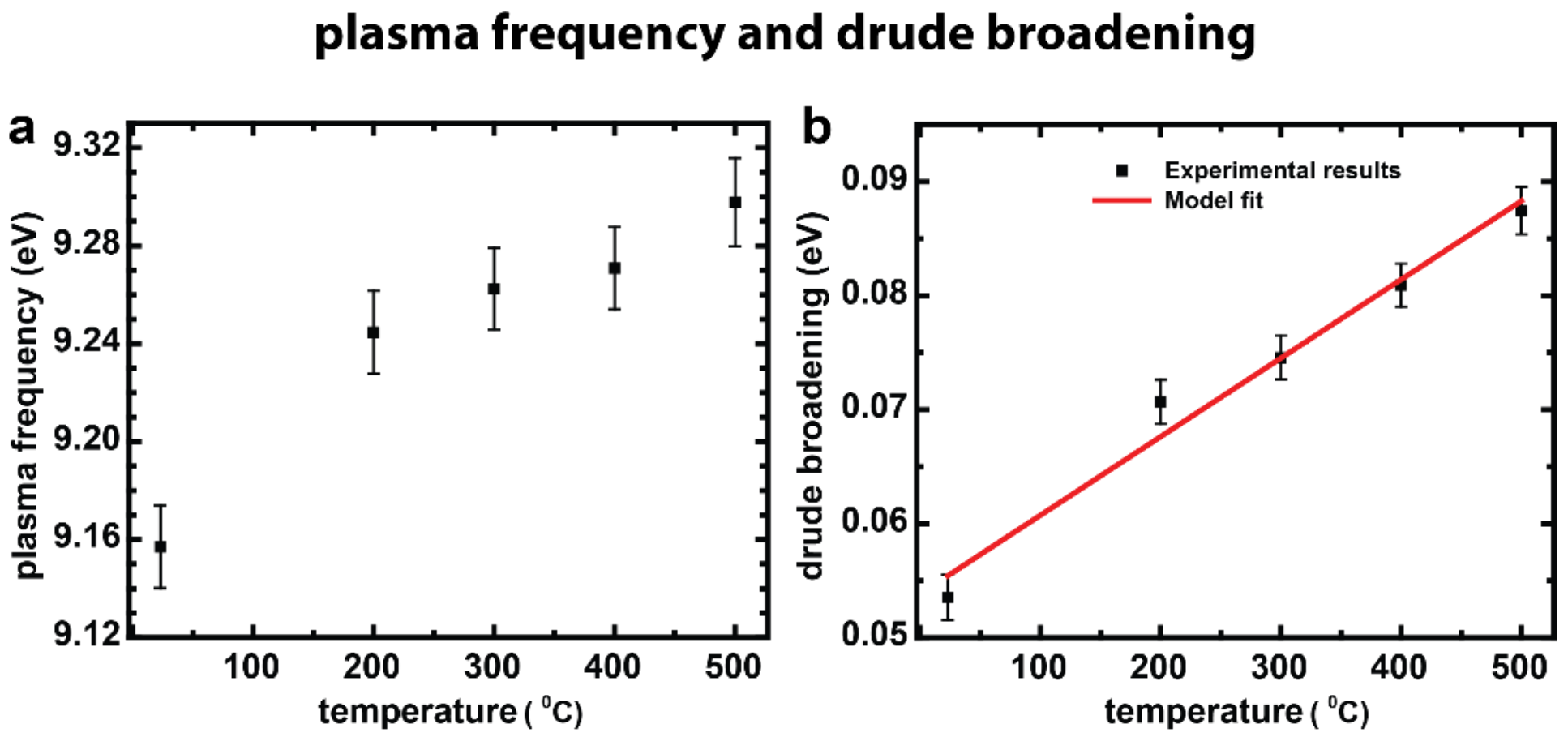

**Figure S8:** Plasma frequency (a) and Drude broadening (b) of the 200-nm-thick single-crystalline gold film. The plasma frequency increase monotonically with increasing temperature unlike the polycrystalline film where it either increases or decreases depending on the temperature range (Figure 6). But the Drude broadening increases with increasing temperature similar to the polycrystalline film. The red curve is the fit obtained using Eqn (1.10) from the manuscript.

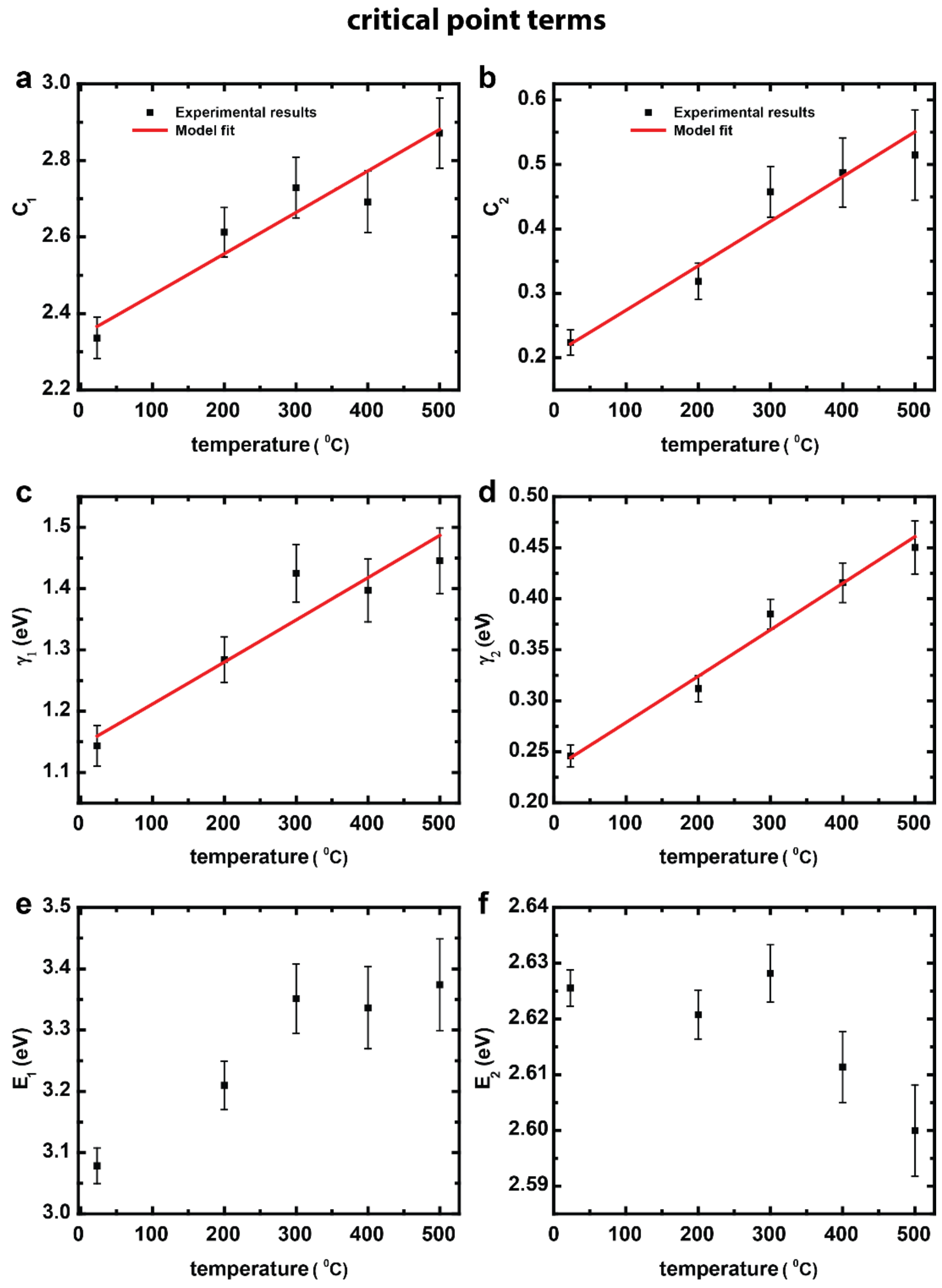

**Figure S9**: Oscillator strengths (a,b), Oscillator dampings (c,d) and Oscillator energies (e,f) of 200-nm-thick single crystalline films. The red curve shows the fit obtained using the empirical expressions discussed in the Theory section of the manuscript.

# Drude and 2 Critical Point Models

The Following form of Drude and 2 Critical Point (DCP) model was used to fit the VASE data for all the samples:

$$\varepsilon(\omega) = \varepsilon_\infty - \frac{\omega_p^2}{\omega^2 + i\Gamma_D\omega} + \sum_{j=1}^{2} C_j\Omega_j \left( \frac{e^{i\phi_j}}{\Omega_j - \omega - i\gamma_j} - \frac{e^{-i\phi_j}}{\Omega_j + \omega + i\gamma_j} \right)$$

The temperature dependent coefficients of the DCP model for different samples are shown below.

## Table S1

## 200-nm-thick poly-crystalline gold film

### First Cycle

| *T* | $\varepsilon_\infty$ | $\omega_{pu}$ (eV) | $\Gamma_D$ (eV) | $C_1$ | $\phi_1$ | $\gamma_1$ (eV) | $E_1$ (eV) | $C_2$ | $\phi_2$ | $\gamma_2$ (eV) | $E_2$ (eV) | MSE |
|---|---|---|---|---|---|---|---|---|---|---|---|---|
| 23 $^0$C | 2.27 | 8.856 | 0.0471 | 2.31 | $-\pi/4$ | 1.215 | 3.082 | 0.226 | $-\pi/4$ | 0.256 | 2.625 | 1.19 |
| 100 $^0$C | 2.45 | 8.863 | 0.0489 | 2.28 | $-\pi/4$ | 1.198 | 3.060 | 0.224 | $-\pi/4$ | 0.273 | 2.620 | 1.04 |
| 200 $^0$C | 2.04 | 9.113 | 0.0496 | 2.64 | $-\pi/4$ | 1.464 | 3.270 | 0.395 | $-\pi/4$ | 0.359 | 2.631 | 0.93 |
| 300 $^0$C | 2.06 | 9.012 | 0.0678 | 2.49 | $-\pi/4$ | 1.357 | 3.218 | 0.371 | $-\pi/4$ | 0.375 | 2.607 | 1.09 |
| 400 $^0$C | 2.03 | 8.978 | 0.0695 | 2.56 | $-\pi/4$ | 1.407 | 3.208 | 0.375 | $-\pi/4$ | 0.398 | 2.599 | 0.98 |
| 500 $^0$C | 1.90 | 8.959 | 0.0815 | 2.54 | $-\pi/4$ | 1.369 | 3.209 | 0.394 | $-\pi/4$ | 0.429 | 2.578 | 0.86 |

# Table S2

## Second Cycle

| $T$ | $\varepsilon_\infty$ | $\omega_{pu}$ (eV) | $\Gamma_D$ (eV) | $C_1$ | $\phi_1$ | $\gamma_1$ (eV) | $E_1$ (eV) | $C_2$ | $\phi_2$ | $\gamma_2$ (eV) | $E_2$ (eV) | MSE |
|---|---|---|---|---|---|---|---|---|---|---|---|---|
| 23 $^0$C | 2.33 | 8.909 | 0.0421 | 2.32 | $-\pi / 4$ | 1.160 | 3.080 | 0.206 | $-\pi / 4$ | 0.235 | 2.620 | 1.41 |
| 100 $^0$C | 2.24 | 8.940 | 0.0460 | 2.41 | $-\pi / 4$ | 1.250 | 3.124 | 0.254 | $-\pi / 4$ | 0.277 | 2.622 | 1.35 |
| 200 $^0$C | 2.23 | 8.985 | 0.0542 | 2.41 | $-\pi / 4$ | 1.313 | 3.171 | 0.259 | $-\pi / 4$ | 0.332 | 2.619 | 1.35 |
| 300 $^0$C | 2.04 | 8.980 | 0.0618 | 2.50 | $-\pi / 4$ | 1.365 | 3.211 | 0.356 | $-\pi / 4$ | 0.362 | 2.608 | 1.10 |
| 400 $^0$C | 1.93 | 8.954 | 0.0720 | 2.47 | $-\pi / 4$ | 1.353 | 3.213 | 0.392 | $-\pi / 4$ | 0.403 | 2.592 | 1.24 |
| 500 $^0$C | 1.66 | 8.865 | 0.0857 | 2.50 | $-\pi / 4$ | 1.359 | 3.233 | 0.405 | $-\pi / 4$ | 0.432 | 2.572 | 0.99 |

## **Table S3**

Third Cycle

| *T* | $\varepsilon_\infty$ | $\omega_{pu}$ (eV) | $\Gamma_D$ (eV) | $C_1$ | $\phi_1$ | $\gamma_1$ (eV) | $E_1$ (eV) | $C_2$ | $\phi_2$ | $\gamma_2$ (eV) | $E_2$ (eV) | MSE |
|---|---|---|---|---|---|---|---|---|---|---|---|---|
| 23 $^0$C | 1.99 | 8.706 | 0.0479 | 2.27 | $-\pi/4$ | 1.205 | 3.110 | 0.225 | $-\pi/4$ | 0.250 | 2.628 | 1.51 |
| 100 $^0$C | 2.41 | 8.834 | 0.0515 | 2.22 | $-\pi/4$ | 1.205 | 3.085 | 0.251 | $-\pi/4$ | 0.281 | 2.622 | 1.34 |
| 200 $^0$C | 2.08 | 8.879 | 0.0578 | 2.36 | $-\pi/4$ | 1.307 | 3.175 | 0.324 | $-\pi/4$ | 0.332 | 2.615 | 0.96 |
| 300 $^0$C | 1.83 | 8.856 | 0.0711 | 2.39 | $-\pi/4$ | 1.301 | 3.194 | 0.349 | $-\pi/4$ | 0.361 | 2.601 | 1.08 |
| 400 $^0$C | 1.85 | 8.847 | 0.0819 | 2.38 | $-\pi/4$ | 1.320 | 3.203 | 0.377 | $-\pi/4$ | 0.398 | 2.590 | 0.98 |
| 500 $^0$C | 1.59 | 8.778 | 0.0893 | 2.48 | $-\pi/4$ | 1.368 | 3.221 | 0.381 | $-\pi/4$ | 0.423 | 2.577 | 0.90 |

# Table S4

# 200-nm-thick single-crystalline gold film

## First Cycle

| $T$ | $\varepsilon_{\infty}$ | $\omega_{pu}$ (eV) | $\Gamma_D$ (eV) | $C_1$ | $\phi_1$ | $\gamma_1$ (eV) | $E_1$ (eV) | $C_2$ | $\phi_2$ | $\gamma_2$ (eV) | $E_2$ (eV) | MSE |
|---|---|---|---|---|---|---|---|---|---|---|---|---|
| 23 $^0$C | 2.64 | 8.991 | 0.0534 | 2.22 | $-\pi/4$ | 1.105 | 3.062 | 0.206 | $-\pi/4$ | 0.235 | 2.618 | 1.41 |
| 200 $^0$C | 2.27 | 9.218 | 0.0725 | 2.57 | $-\pi/4$ | 1.337 | 3.258 | 0.371 | $-\pi/4$ | 0.334 | 2.629 | 1.00 |
| 300 $^0$C | 2.52 | 9.310 | 0.0712 | 2.45 | $-\pi/4$ | 1.337 | 3.277 | 0.465 | $-\pi/4$ | 0.391 | 2.623 | 0.96 |
| 400 $^0$C | 2.13 | 9.324 | 0.0805 | 2.63 | $-\pi/4$ | 1.405 | 3.352 | 0.517 | $-\pi/4$ | 0.426 | 2.612 | 0.93 |
| 500 $^0$C | 1.79 | 9.323 | 0.0898 | 2.73 | $-\pi/4$ | 1.425 | 3.400 | 0.579 | $-\pi/4$ | 0.469 | 2.602 | 0.91 |

# Table S5

## Second Cycle

| $T$ | $\varepsilon_\infty$ | $\omega_{pu}$ (eV) | $\Gamma_D$ (eV) | $C_1$ | $\phi_1$ | $\gamma_1$ (eV) | $E_1$ (eV) | $C_2$ | $\phi_2$ | $\gamma_2$ (eV) | $E_2$ (eV) | MSE |
|---|---|---|---|---|---|---|---|---|---|---|---|---|
| 23 $^0$C | 2.82 | 9.157 | 0.0535 | 2.34 | $-\pi/4$ | 1.144 | 3.079 | 0.224 | $-\pi/4$ | 0.246 | 2.626 | 1.40 |
| 200 $^0$C | 2.33 | 9.245 | 0.0707 | 2.61 | $-\pi/4$ | 1.284 | 3.210 | 0.319 | $-\pi/4$ | 0.312 | 2.621 | 0.95 |
| 300 $^0$C | 1.91 | 9.262 | 0.0746 | 2.73 | $-\pi/4$ | 1.425 | 3.351 | 0.458 | $-\pi/4$ | 0.384 | 2.628 | 0.97 |
| 400 $^0$C | 1.92 | 9.271 | 0.0809 | 2.69 | $-\pi/4$ | 1.397 | 3.336 | 0.487 | $-\pi/4$ | 0.416 | 2.611 | 0.87 |
| 500 $^0$C | 1.67 | 9.298 | 0.0875 | 2.87 | $-\pi/4$ | 1.445 | 3.374 | 0.514 | $-\pi/4$ | 0.450 | 2.599 | 0.89 |

# Table S6

## 50-nm-thick gold film

| $T$ | $\varepsilon_\infty$ | $\omega_{pu}$ (eV) | $\Gamma_D$ (eV) | $C_1$ | $\phi_1$ | $\gamma_1$ (eV) | $E_1$ (eV) | $C_2$ | $\phi_2$ | $\gamma_2$ (eV) | $E_2$ (eV) | MSE |
|---|---|---|---|---|---|---|---|---|---|---|---|---|
| 23 $^0$C | 1.90 | 8.800 | 0.0752 | 2.54 | $-\pi/4$ | 1.344 | 3.112 | 0.245 | $-\pi/4$ | 0.272 | 2.641 | 1.25 |
| 100 $^0$C | 2.05 | 8.876 | 0.0752 | 2.57 | $-\pi/4$ | 1.350 | 3.106 | 0.250 | $-\pi/4$ | 0.281 | 2.633 | 1.23 |
| 200 $^0$C | 1.85 | 8.927 | 0.0771 | 2.67 | $-\pi/4$ | 1.444 | 3.194 | 0.327 | $-\pi/4$ | 0.339 | 2.631 | 0.98 |
| 250 $^0$C | 1.59 | 8.713 | 0.0706 | 2.54 | $-\pi/4$ | 1.436 | 3.193 | 0.335 | $-\pi/4$ | 0.356 | 2.622 | 0.97 |
| 300 $^0$C | 1.43 | 8.449 | 0.0725 | 2.25 | $-\pi/4$ | 1.464 | 3.242 | 0.406 | $-\pi/4$ | 0.404 | 2.614 | 0.90 |
| 350 $^0$C | 1.38 | 8.365 | 0.0726 | 2.24 | $-\pi/4$ | 1.460 | 3.221 | 0.390 | $-\pi/4$ | 0.408 | 2.606 | 0.86 |
| 400 $^0$C | 2.34 | 8.683 | 0.0877 | 2.26 | $-\pi/4$ | 1.286 | 3.046 | 0.289 | $-\pi/4$ | 0.381 | 2.589 | 0.85 |
| 450 $^0$C | 2.46 | 9.002 | 0.1451 | 2.45 | $-\pi/4$ | 1.311 | 3.140 | 0.346 | $-\pi/4$ | 0.408 | 2.594 | 0.79 |
| 500 $^0$C | 2.39 | 8.949 | 0.1770 | 2.18 | $-\pi/4$ | 1.270 | 3.226 | 0.488 | $-\pi/4$ | 0.479 | 2.585 | 0.85 |

# Table S7

# 30-nm-thick gold film

| $T$ | $\varepsilon_\infty$ | $\omega_{pu}$ (eV) | $\Gamma_D$ (eV) | $C_1$ | $\phi_1$ | $\gamma_1$ (eV) | $E_1$ (eV) | $C_2$ | $\phi_2$ | $\gamma_2$ (eV) | $E_2$ (eV) | MSE |
|---|---|---|---|---|---|---|---|---|---|---|---|---|
| 23 $^0$C | 1.91 | 8.670 | 0.0740 | 2.57 | $-\pi/4$ | 1.318 | 3.078 | 0.195 | $-\pi/4$ | 0.267 | 2.634 | 1.10 |
| 100 $^0$C | 1.22 | 8.724 | 0.0769 | 2.98 | $-\pi/4$ | 1.899 | 3.434 | 0.440 | $-\pi/4$ | 0.392 | 2.665 | 1.09 |
| 200 $^0$C | 1.43 | 8.767 | 0.0780 | 2.91 | $-\pi/4$ | 1.619 | 3.283 | 0.353 | $-\pi/4$ | 0.376 | 2.642 | 1.10 |
| 250 $^0$C | 1.21 | 8.629 | 0.0715 | 2.68 | $-\pi/4$ | 1.674 | 3.381 | 0.463 | $-\pi/4$ | 0.429 | 2.634 | 1.08 |
| 300 $^0$C | 1.61 | 8.525 | 0.0744 | 2.26 | $-\pi/4$ | 1.448 | 3.239 | 0.451 | $-\pi/4$ | 0.442 | 2.615 | 1.04 |
| 350 $^0$C | 2.12 | 8.547 | 0.0890 | 2.08 | $-\pi/4$ | 1.302 | 3.128 | 0.399 | $-\pi/4$ | 0.436 | 2.594 | 0.97 |
| 400 $^0$C | 2.64 | 8.773 | 0.152 | 2.10 | $-\pi/4$ | 1.263 | 3.131 | 0.404 | $-\pi/4$ | 0.451 | 2.598 | 0.99 |
| 450 $^0$C | 3.69 | 8.773 | 0.354 | 0.68 | $-\pi/4$ | 0.607 | 3.401 | 0.977 | $-\pi/4$ | 0.568 | 2.621 | 1.41 |